\documentclass[acmsmall]{acmart}

\usepackage[linesnumbered,ruled,vlined]{algorithm2e}

\usepackage[export]{adjustbox} 
\usepackage[table]{xcolor}
\usepackage{balance}
\usepackage[utf8]{inputenc}
\usepackage[framemethod=TikZ]{mdframed}
\usepackage{tcolorbox}
\usepackage{graphicx}
\usepackage{float}
\usepackage{array}
\usepackage{tabularx}
\usepackage{multirow}
\usepackage{amsmath}
\usepackage{amssymb}
\usepackage{threeparttable}
\usepackage{setspace}
\usepackage{color}
\usepackage{colortbl}
\usepackage{xcolor}
\usepackage{soul}
\usepackage{booktabs}
\usepackage{makecell}
\usepackage{hyperref}
\usepackage{enumitem}
\usepackage{listings}
\usepackage{verbatim}
\usepackage{tikz}
\usepackage{arydshln}
\usepackage{subcaption}
\usepackage{parselines} 
\usepackage{wrapfig}
\usepackage{url}
\usepackage{pifont}

\renewcommand{\arraystretch}{1.2} 

\usepackage[skip=1pt,labelfont=bf]{caption}
\usepackage{framed} 

\newcommand*\circled[1]{\tikz[baseline=(char.base)]{
            \node[shape=circle,draw,inner sep=0.75pt] (char) {#1};}}

\newcommand{\tool}{IaCGen}

\definecolor{customblue}{HTML}{006ca6}
\definecolor{customgreen}{HTML}{009264}
\definecolor{custombrown}{HTML}{ff3d00}
\AtEndPreamble{
 \usepackage{hyperref}
 \hypersetup{
  colorlinks = true,
  linkcolor = customblue,
  anchorcolor = customblue,
  citecolor = customgreen,
  filecolor = customblue,
  urlcolor = customblue
 }
}

\AtBeginDocument{%
  \providecommand\BibTeX{{%
    \normalfont B\kern-0.5em{\scshape i\kern-0.25em b}\kern-0.8em\TeX}}}

\newcommand{\ty}[1]{\textcolor{blue}{#1}}

\newcommand{\find}[1]{
\begin{tcolorbox}[leftrule=1mm,toprule=0mm,bottomrule=0mm,left=1pt,right=2pt,top=2pt,bottom=2pt
]
\em #1
\end{tcolorbox}
}

\newcommand{\paratitle}[1]{
    \setlength{\parskip}{0.3\baselineskip}  \noindent\underline{\textbf{#1}.}%
}

\def\BibTeX{{\rm B\kern-.05em{\sc i\kern-.025em b}\kern-.08em
    T\kern-.1667em\lower.7ex\hbox{E}\kern-.125emX}}

\AtBeginDocument{%
  \providecommand\BibTeX{{%
    Bib\TeX}}}

\setcopyright{acmlicensed}
\copyrightyear{2025}
\acmYear{2025}
\acmDOI{XXXXXXX.XXXXXXX}
\acmConference[FSE]{}{2026}{Montreal, Canada}
\acmISBN{978-1-4503-XXXX-X/YY/MM}

\begin{document}

\author{Tianyi Zhang}
\authornote{Both authors contributed equally to this research.}
\affiliation{%
  \institution{Australian National University}
  \city{Canberra}
  \country{Australia}
}
\email{tianyi.zhang@anu.edu.au}

\author{Shidong Pan}
\authornotemark[1]
\email{shidong.pan@nyu.edu}
\affiliation{%
  \institution{New York University \& Columbia University}
  \country{USA}
}

\author{Zejun Zhang}
\email{zejun.zhang@ntu.edu.sg}
\affiliation{%
  \institution{Nanyang Technological University}
  \country{Singapore}
}

\author{Zhenchang Xing}
\email{zhenchang.xing@data61.csiro.au}
\affiliation{%
 \institution{CSIRO's Data61}
 \country{Australia}
}

\author{Xiaoyu Sun}
\authornote{Corresponding author.}
\email{xiaoyu.sun1@anu.edu.au}
\affiliation{%
  \institution{Australian National University}
  \country{Australia}
}




\title{Deployability-Centric Infrastructure-as-Code Generation: Fail, Learn, Refine, and Succeed through LLM-Empowered DevOps Simulation}

\begin{abstract}
Infrastructure-as-Code (IaC) generation holds significant promise for automating cloud infrastructure provisioning. 
Recent advances in Large Language Models (LLMs) present a promising opportunity to democratize IaC development by generating deployable infrastructure templates from natural language descriptions.
However, current evaluation focuses on syntactic correctness while ignoring deployability, the critical measure of the utility of IaC configuration files. 
Six state-of-the-art LLMs performed poorly on deployability, achieving only 20.8$\sim$30.2\% deployment success rate on the first attempt.
In this paper, we construct DPIaC-Eval, the first deployability-centric IaC template benchmark consisting of 153 real-world scenarios cross 58 unique services.
Also, we propose an LLM-based deployability-centric framework, dubbed IaCGen, that uses iterative feedback mechanism encompassing format verification, syntax checking, and live deployment stages, thereby closely mirroring the real DevOps workflows. 
Results show that IaCGen can make 54.6$\sim$91.6\% generated IaC templates from all evaluated models deployable in the first 10 iterations.
Additionally, human-in-the-loop feedback that provide direct guidance for the deployability errors, can further boost the performance to over 90\% passItr@25 on all evaluated LLMs.
Furthermore, we explore the trustworthiness of the generated IaC templates on user intent alignment and security compliance.
The poor performance (25.2\% user requirement coverage and 8.4\% security compliance rate) indicates a critical need for continued research in this domain. 
\end{abstract}



\maketitle

\section{Introduction}
The digital transformation of enterprises has fundamentally altered how organizations manage and deploy computing infrastructure. As cloud adoption accelerates globally, with over 82\% of enterprises now utilizing cloud services~\cite{rackspace-report}, the complexity and scale of infrastructure management have increased exponentially. Infrastructure-as-Code (IaC) has emerged as the default standard for managing this complexity, enabling organizations to define, provision, and maintain cloud resources through programmable templates that ensure consistency, repeatability, and version control across deployments~\cite{kumara2021s, chinamanagonda2019automating, sokolowski2022infrastructure}. This paradigm shift from manual infrastructure management to code-driven automation has become essential for modern DevOps practices~\cite{basher2019devops}, supporting the rapid deployment cycles and scalability requirements of contemporary software systems.

However, the widespread adoption of IaC faces significant barriers rooted in its inherent technical complexity and the specialized expertise it demands~\cite{cusumano2010cloud, qiu2023simplifying, iac_low}. Creating IaC templates requires a deep understanding of cloud service architectures, resource interdependencies, and platform-specific constraints that extend far beyond general programming knowledge~\cite{basher2019devops, li2024incremental}. The steep learning curve of IaC languages creates substantial obstacles for development teams lacking dedicated DevOps expertise~\cite{kon2024iac, hashicorp2023}. This expertise gap not only slows the development speed but also increases the risk of misconfigurations that lead to security vulnerabilities and service outages~\cite{nasiri2024towards, drosos2024your}.

Recent advances in Large Language Models (LLMs) present a promising opportunity to democratize IaC development by generating deployable infrastructure templates from natural language descriptions~\cite{chen2021evaluating, liu2023your, liao2024code, liu2025ai, liao2025navigating, zhou2025declarui}. An effective LLM-based IaC generator could potentially transform infrastructure automation by enabling developers to describe their infrastructure requirements in natural language and receive working deployable IaC in return. This capability would significantly accelerate development workflows in DevOps and reduce the expertise barrier for cloud adoption.


However, previous research indicates that LLMs demonstrate considerably lower performance in IaC generation compared to general programming tasks~\cite{kon2024iac, ragothaman2024optimizing, han2024chase}, with initial success rates as low as 19\%.
We notice that existing research~\cite{kon2024iac, ragothaman2024optimizing, palavalli2024using} suffers from a critical limitation: an overemphasis on syntactic correctness while neglecting the fundamental requirement of deployability. In operational contexts, deployability represents the primary measure of IaC template quality, as non-deployable templates provide no practical value regardless of syntactic correctness. 


To mitigate research gaps, we present \textbf{DPIaC-Eval, a novel benchmark comprising 153 real-world infrastructure tasks across 58 unique AWS services}, with expert-authored natural language requirements and verified deployable IaC template as reference templates. 
The formative study shows that mainstream LLMs can only achieve a merely 50\% deployment success rate even in the simplest scenarios (with only one resource and one parameter) in its first attempt.
This poor performance stems from fundamental challenges in IaC generation: unlike traditional programming where logic errors are often apparent~\cite{sun2021taming}, IaC's declarative nature obscures issues until execution, requiring iterative testing and validation~\cite{sun2023lazycow,sun2023taming,sun2022mining}. 
Therefore, we propose \textbf{\tool{}, a novel deployability-centric IaC generation framework}.
\tool{} is integrated through an iterative feedback mechanism across three deployability-oriented validation stages, format verification, syntax checking, and live deployment, which mirrors the real DevOps workflows. 
Such design enables models to learn from syntactic and deployment failures, progressively refining templates.

We then evaluate the effectiveness of IaCGen with six popular LLMs. 
Results show that using LLM alone achieves only 20.8$\sim$30.2\% deployment success rate, 
while IaCGen achieves 39.0\%$\sim$66.7\% in five iterations and 54.6\%$\sim$91.6\% when the number of iterations is doubled. 
We further summarize the top-5 common errors, including \textit{Missing Value}, \textit{Self-defined Property}, \textit{Null Substitution}, \textit{Unnecessary Whitespace}, and \textit{Arbitrary Default Value}.
Results reveal that different models tend to make different types of error, with the Claude models showing the best performance in avoiding all types of errors. 
Additionally, we evaluate whether IaCGen can achieve better performance on generating IaC templates with human-in-the-loop feedback. 
We find that human feedback, which provide direct guidance for the error reflected from the error message, can improve overall deployment success rate by over 19.5\%, help all six LLMs achieve more than 90\% passItr@25. 

We further explore the trustworthiness of the generated IaC templates. 
First, we evaluate the \underline{user intent matching} on \textit{resources} and \textit{attributes} against the 
reference template 
and the 25.2\% average coverage indicates critical challenges remain in user intent alignment.
Additionally, we utilize Checkov to evaluate the \underline{security compliance}.
The low compliance rate (on applicable policies) of 8.4\% indicates a critical need for continued research in this domain.
Furthermore, we discuss typical security policy violations (70 distinct security policy violations cross all generated IaC templates), automation potential with human-in-the-loop performance, the implications of DevOps-inspired LLM-based framework, and the threats to validity.

In summary, our paper makes the following contributions:

\begin{itemize}[leftmargin=2em]
    \item DPIaC-Eval, the first deployability-centric benchmark for IaC template generation, containing 153 real-world IaC tasks across five difficulty levels and 58 AWS services, with multi-dimensional validation spanning syntax, deployment, user intent, and security.
    \item \tool{}, an LLM-based framework on IaC template generation with iterative feedback mechanism, achieving the state-of-the-art performance. 
    \item Empirical evidences about model performance across multiple dimensions of IaC template quality, providing insights into common failure patterns and priority areas for improvement.
\end{itemize}
\section{Background}

\subsection{Infrastructure-as-Code}
IaC represents a paradigm for managing cloud infrastructure through declarative configuration files, enabling automated provisioning and management of cloud resources. The terms ``\textit{IaC}'', ``\textit{IaC language}'', and ``\textit{IaC tool}'' are used interchangeably in the literature to refer to the domain-specific languages designed for infrastructure specification and deployment. Prominent examples include AWS CloudFormation, HashiCorp Terraform, and Azure Resource Manager. Each IaC language provides its own syntax and capabilities for defining infrastructure configurations. \textbf{IaC templates} are the configuration files that contain the infrastructure specifications written in a particular IaC language. These templates are typically expressed in declarative formats such as YAML, JSON, or HashiCorp Configuration Language, depending on the target IaC language.

\begin{figure}[t] 
    \setlength{\belowcaptionskip}{-13pt}
    \centering
    
    \begin{subfigure}[t]{0.48\textwidth}
        \setlength{\abovecaptionskip}{1.8pt}
        \setlength{\belowcaptionskip}{1pt}
        \centering
        \includegraphics[width=\linewidth, valign=t]{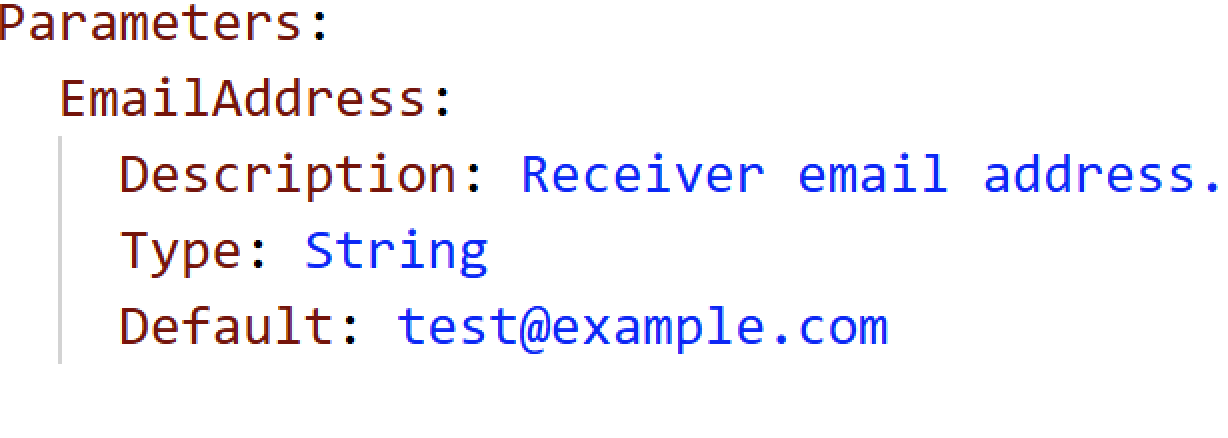}
        \caption{Parameters section}
        \label{fig:image1}
    \end{subfigure}
    \hfill
    \begin{subfigure}[t]{0.48\textwidth}
        \setlength{\abovecaptionskip}{3.5pt}
        \setlength{\belowcaptionskip}{1pt}
        \centering
        \includegraphics[width=\linewidth, valign=t]{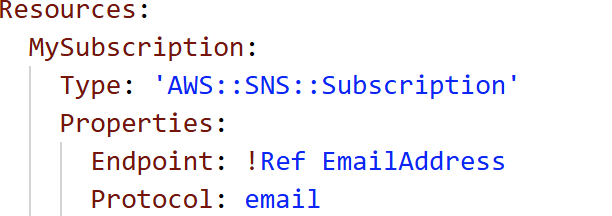}
        \caption{Resources section}
        \label{fig:image2}
    \end{subfigure}
    
    \caption{A sample CloudFormation template creating an SNS email subscription with notification recipient email address defined with parameter.}
    \label{fig:cf_example}
\end{figure}

IaC templates are structured around two fundamental components (resource and parameter). \textbf{Resources} represent the cloud services and infrastructure components to be provisioned. The term ``resources'' is used instead of ``services'' because a single cloud service may expose multiple resource types. For instance, Amazon SNS, a message delivery service, provides four distinct resource types (Topic, Subscription, TopicPolicy, and TopicInlinePolicy). In AWS CloudFormation, resources follow the hierarchical naming convention ``\textit{AWS::ServiceName::ResourceType}'', clearly delineating the service provider, service name, and specific resource type. \textbf{Parameters} enable template customization by allowing external specification of resource property values. Parameters improve template reusability and flexibility by enabling the same template to be deployed across different environments or use cases with varying configurations. Without parameters, templates would be static and require modification for each deployment scenario. \textbf{Resource properties} define the specific attributes and configurations of infrastructure components, analogous to class attributes in object-oriented programming. For example, as shown in Fig.~\ref{fig:cf_example}, an ``\textit{AWS::SNS::Subscription}'' resource includes properties such as ``\textit{Endpoint}'' (specifying the notification destination) and ``\textit{Protocol}'' (defining the delivery method). These properties determine the runtime behavior and integration patterns of the provisioned resources. The knowledge barrier and widespread adoption of IaC underscore the need for automated IaC template generation methods to effectively support developers.

    

    

\subsection{Difference Between IaC Tools}\label{sec:difference-between-IaC-tools}
Terraform (TF) and CloudFormation (CF) represent two distinct types of IaC tools, each with unique format language, design philosophy, and operational characteristics.
TF uses the HashiCorp Configuration Language (HCL), a declarative language designed with a provider-based architecture that enables multi-cloud deployments across AWS, Azure, Google Cloud, and numerous other platforms~\cite{pum2024cloudformation}, while CF uses JSON or YAML to define resources in a hierarchical structure with deep integration into the AWS's service ecosystem~\cite{hash-tf-cf-compare}. 
Additionally, CF focuses exclusively on AWS, providing native integration with AWS services, security models, and deployment patterns, including features like AWS-specific intrinsic functions and cross-stack references~\cite{geeks-tf-cf-compare}.

In addition to the syntactic validity, the program deployability fatally determines real-world IaC template utility.
TF's stateful architecture requires explicit state management and manual resource cleanup procedures~\cite{geeks-tf-cf-compare}, making systematic deployability evaluation across hundreds of templates impractical. 
Also, TF's lack of native rollback mechanisms means that failed deployments can leave partially created resources in indeterminate states, requiring manual cloud resources cleaning~\cite{tf-limitation-first, tf-limitation-second, hash-tf-cf-compare}. These operational constraints might explain why existing TF-based benchmark limit evaluation to static analysis~\cite{kon2024iac}, leaving the crucial deployability dimension unaddressed.

CF, as AWS's native IaC language, provides unique methodological advantages that make comprehensive deployability testing feasible. 
When deploying a CF template, AWS automatically creates a stack to hold the collection of resources defined in the template~\cite{cf-stack}. 
Users can create, update, and delete entire resource collections by manipulating stacks, enabling systematic removal of all created resources and restoration of the cloud environment to its initial state through simple stack deletion~\cite{cf-stack}. 
Therefore, to explore the automated IaC template generation especially for deployability, we focus on the CloudFormation in this paper.



%
%

%
%

\section{Benchmark: DPIaC-Eval}\label{sec_benchmark}
Most existing research in code generation has predominantly concentrated on general-purpose programming languages~\cite{liao2024code,gu2023llm,koziolek2024llm,coignion2024performance}, often overlooking the distinct challenges inherent to IaC. Unlike conventional software programs, IaC templates must not only be syntactically correct but also semantically valid and deployable within specific cloud environments. 
Thus, to systematically evaluate the effectiveness of IaC template generation, we develop the first \underline{\textbf{d}}e\underline{\textbf{p}}loyability-focused benchmark, DPIaC-Eval, comprising 153 diverse CloudFormation IaC templates.


\subsection{Benchmark Construction}
\begin{figure}[t] 
    \setlength{\abovecaptionskip}{8pt}
    \setlength{\belowcaptionskip}{-10pt}
    \centering
    \includegraphics[width=0.95\linewidth]{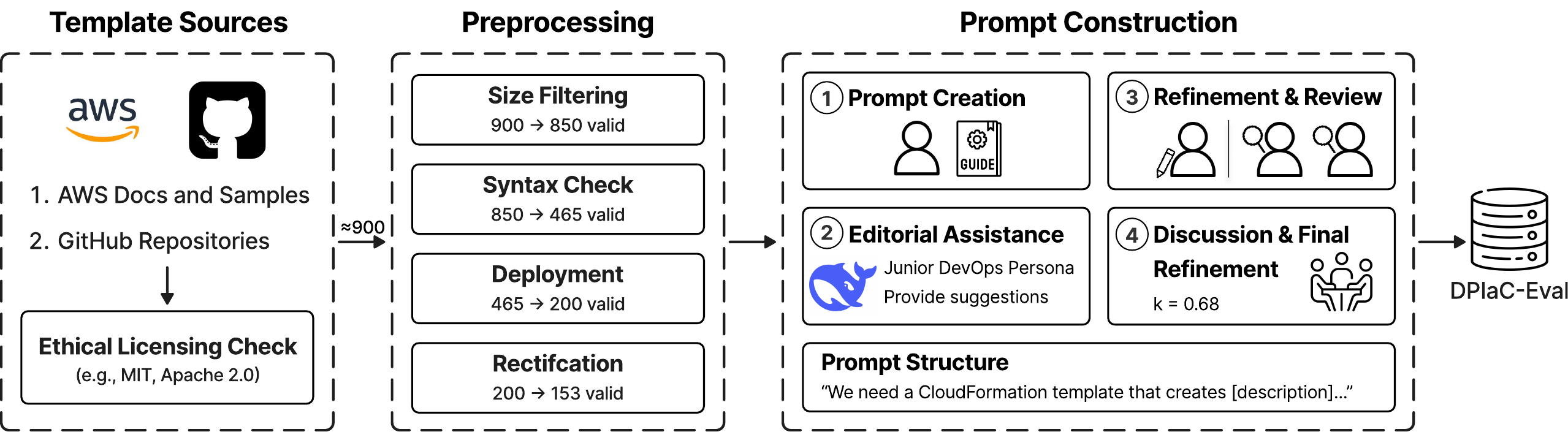}
    \captionsetup{justification=centering}
    \caption{Workflow of benchmark DPIaC-Eval construction.} %
    \label{fig:dataset_construction}
\end{figure}

The benchmark construction process (Fig.~\ref{fig:dataset_construction}) follows a systematic and reproducible workflow designed to collect high-quality AWS CloudFormation templates and ensure adequacy. 

\paratitle{Template Sources}~Templates are sourced from two primary channels to ensure diversity and quality: (1) official AWS documentation and sample libraries, including the AWS CloudFormation Sample Templates repository~\cite{aws_cfn_templates}; (2) GitHub repositories that contain CloudFormation template which we find within the AWS Samples GitHub organization~\cite{aws_samples_github} and a dataset containing URLs of repositories that use CloudFormation templates~\cite{cirlan2024mining}. Ethical considerations are enforced by confirming appropriate licensing (e.g., MIT, Apache 2.0) before inclusion, ensuring that all templates are legally permissible for research purposes.

\paratitle{Preprocessing}~
Following an initial collection of approximately 900 templates, candidates undergo automated preprocessing through a multistage filtering funnel. First, size filtering removes templates that exceed 8,000 tokens or 600 lines, ensuring compatibility with the common LLM generation capacities, resulting in 850 templates.
Next, syntax validation verifies templates for structural correctness using cfn-linter~\cite{cfn-linter}, yielding 465 valid candidates. Templates then undergo deployment testing using the AWS SDK (boto3) with a sandbox AWS account configured with least-privilege IAM roles, reducing the dataset to 200 templates. Finally, a rectification process addresses common deployment failures, including cross-stack dependencies, parameter value completion, missing VPC resources, outdated value references, and dependency ordering adjustments. After rectification, our final dataset comprises 153 high-quality, deployable CloudFormation templates that serve as the foundation for our subsequent IaC generation experiments.


\paratitle{Prompt Construction}~
Although CloudFormation templates include a Description section, our analysis of over 900 templates revealed that most omit this section or provide inadequate summaries (e.g., ``\textit{Deploy Single EC2 Linux Instance}'' for a 170-line template with complex configurations). This necessitated the manual creation of reasonable natural language descriptions. 

For each template, one DevOps practitioner with three years of AWS and CloudFormation experience authored natural language prompts describing the template's intended functionality. All prompts follow a standardized structure: ``\textit{We need a CloudFormation template that creates [description of purpose and required services].}'' The practitioner adhered to specific guidelines to ensure prompt quality: (1) capture essential infrastructure requirements without omitting critical dependencies, (2) reflect realistic practitioner requests that balance specificity with implementation flexibility, and (3) avoid overly prescriptive implementation details.

To enhance prompt quality, we employ DeepSeek-R1 as an editorial assistant, instructed to act as a junior DevOps engineer reviewing the human-authored descriptions alongside original templates. 
The AI assistant helped refine grammar, identify ambiguities, and suggest missing descriptions. For instance, it identified when prompts omitted critical networking dependencies or contained unclear resource relationships. The original practitioner reviewed all AI suggestions and refined prompts as necessary. For validation, two additional DevOps practitioners with equivalent expertise independently reviewed all prompts. The three practitioners discussed discrepancies and collaboratively finalized the descriptions through consensus. 
This rigorous validation process achieved an initial inter-rater agreement of 0.68 (Cohen's Kappa), indicating substantial agreement according to established interpretation guidelines. The resulting prompts remain human-authored and focus on high-level functionality without implementation specifics, mirroring how developers typically specify infrastructure requirements in practice.



%
%

%
%

\begin{figure}[t] 
    \setlength{\belowcaptionskip}{-15pt}
    \centering
    
    \begin{subfigure}[t]{0.63\textwidth}
        \setlength{\abovecaptionskip}{1pt}
        \setlength{\belowcaptionskip}{2pt}
        \centering
        \includegraphics[width=\linewidth, valign=t]{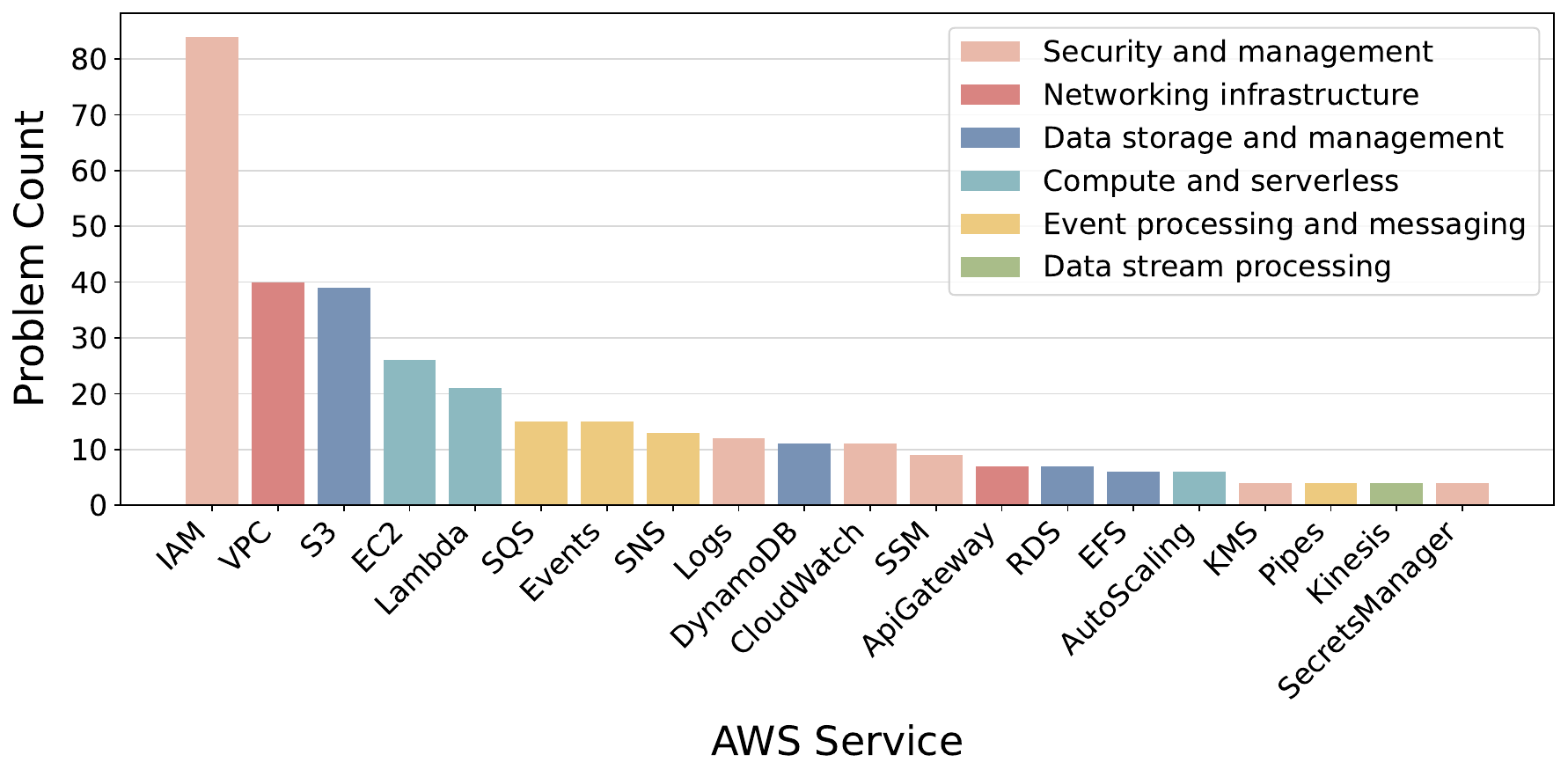}
        \caption{AWS service distribution}
        \label{fig:dataset_services}
    \end{subfigure}
    \hfill
    \begin{subfigure}[t]{0.33\textwidth}
        \setlength{\abovecaptionskip}{9pt}   
        \setlength{\belowcaptionskip}{2pt}
        \centering
        \includegraphics[width=\linewidth, valign=t]{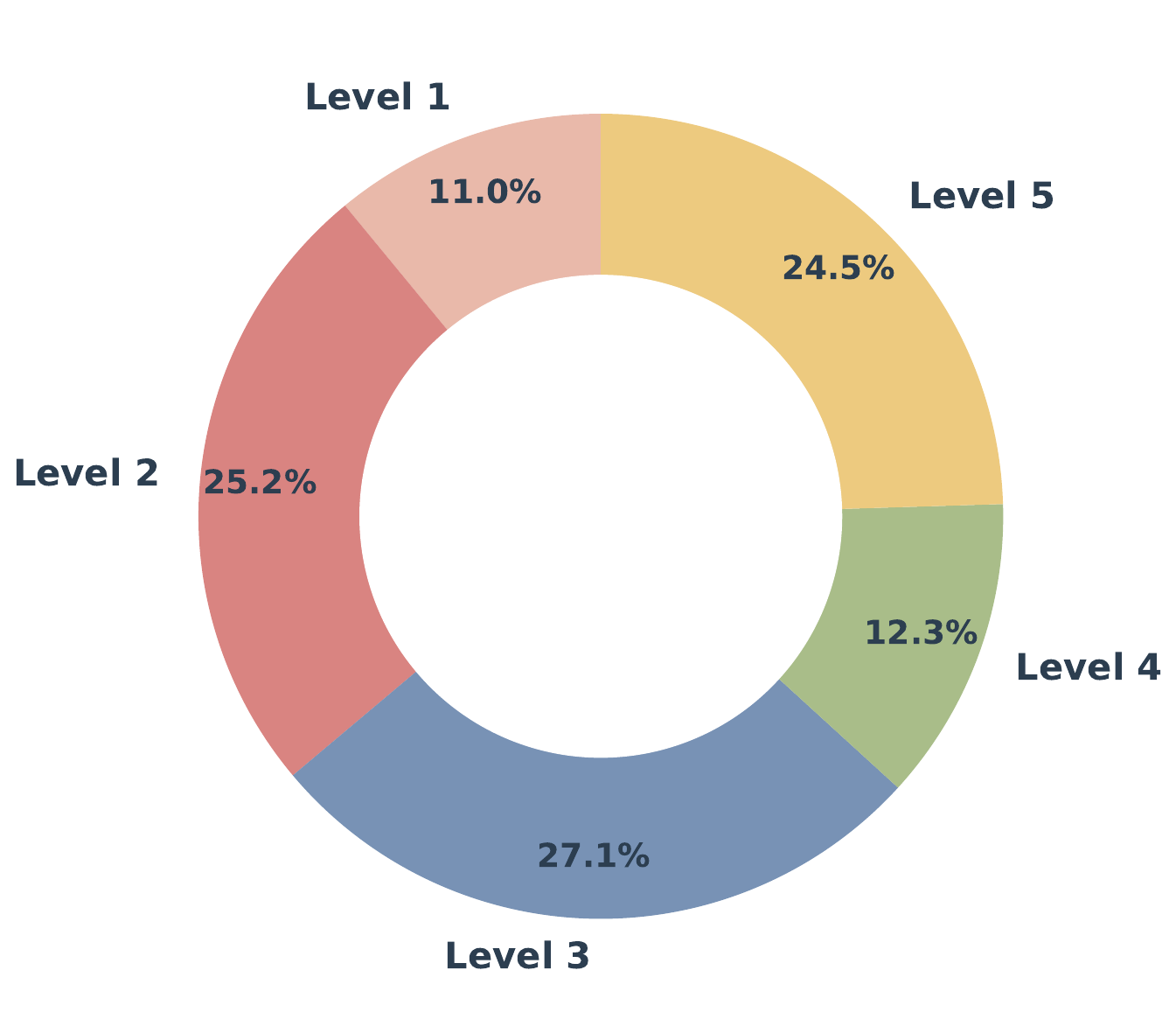}
        \caption{Difficulty level distribution}
        \label{fig:dataset_difficulty_levels}
    \end{subfigure}
    
    \caption{Benchmark characteristics. (a) Distribution of AWS services, presenting the top 20 out of 58 most frequent services. The full list is available in the replication package. (b) Distribution of IaC template difficulty levels, showing balanced coverage across five levels. } 
    \label{fig:benchmark_info}
\end{figure}

\subsection{Benchmark Characteristics}

Each entry in the benchmark contains two essential elements: (1) a natural language prompt describing the infrastructure requirements and (2) the 
reference 
CloudFormation template file path. 
The benchmark comprises 58 commonly used AWS cloud infrastructure services~\cite{drosos2024your, jain2023skyplane, stoica2021cloud}. 
Fig.~\ref{fig:dataset_services} illustrates the 20 most frequent services in our dataset, with frequency represented by the number of templates in which each service appears. The distribution spans the full stack of infrastructure components typically required in modern cloud environments: (1) Compute and serverless: EC2, Lambda, and AutoScaling; (2) Data storage and management: S3, DynamoDB, RDS, and Elastic File System (EFS); (3) Networking infrastructure: Virtual Private Cloud (VPC) and API Gateway; (4) Security and management: Identity and Access Management (IAM), CloudWatch, Logs, and Systems Manager (SSM), Key Management Service (KMS), and SecretsManager; (5) Event processing and messaging: EventBridge (Events), Simple Queue Service (SQS), Pipes, and Simple Notification Service (SNS); (6) Data stream processing: Kinesis. This comprehensive coverage ensures our benchmark effectively evaluates LLMs' capabilities across the diverse range of services essential for real-world cloud infrastructure deployment. The service distribution shown in 
\begin{wraptable}{l}{0.48\linewidth} 
\vspace{-8pt} 

\caption{The criteria of difficulty levels 1-5.}
\vspace{2pt}   
\centering
\label{tab:dataset}
\resizebox{\linewidth}{!}{%
\begin{tabular}{cccc}
\hline
\rowcolor[HTML]{DADADA} 
\textbf{Difficulty Level} & \textbf{LoC} & \textbf{\# Resource} & \textbf{\# Parameter} \\
\hline
\rowcolor[HTML]{F4F4F4} 
1  & $<$~50 & $<$~2 & $<$~2 \\
\rowcolor[HTML]{DADADA} 
2  & $<$~100 & $<$~4 & $<$~5 \\
\rowcolor[HTML]{F4F4F4} 
3  & $<$~150 & $<$~6 & $<$~9 \\
\rowcolor[HTML]{DADADA} 
4  & $<$~200 & $<$~12 & $<$~14 \\
\rowcolor[HTML]{F4F4F4} 
5  & $\geq 200$ & $\geq 12$ & $\geq 14$ \\
\hline
\end{tabular}
}%
\vspace{-7pt} 
\end{wraptable}

Fig.~\ref{fig:dataset_services} reflects real-world usage patterns in production environments, with foundational services like IAM and VPC appearing with higher frequency because they provide the security and networking underpinnings for most cloud architectures. Specialized services demonstrate a proportionally lower representation, mirroring their more targeted usage in production systems.

To facilitate systematic evaluation, we classify templates into five difficulty levels, building upon the design established by IaC-Eval~\cite{kon2024iac}. 
Acknowledging the inherently subjective nature of drawing boundaries for different difficulty levels, the classification is based on objective metrics including the line of code (LoC) in the IaC template, the number of resources required, and the number of configured parameters. The criteria of difficulty level (1-5) are shown in Table~\ref{tab:dataset}.
As shown in Fig.~\ref{fig:dataset_difficulty_levels}, the benchmark demonstrates a balanced distribution across difficulty levels.
This distribution and range ensure the benchmark effectively challenges automated generation methods in various IaC template scenarios while maintaining real-world relevance.

\subsection{Comparison with IaC-Eval Benchmark}
Both DPIaC-Eval and IaC-Eval~\cite{kon2024iac} cover around 60 AWS services, though each includes unique services not found in the other.
DPIaC-Eval presents greater challenges than existing IaC benchmarks, with longer average line of code (155 vs. 42) and more resources per template (7 vs. 4). 
More importantly, DPIaC-Eval provides comprehensive coverage in four critical dimensions: syntax correctness, deployability, user intent matching, and security. 
In contrast, IaC-Eval supports only syntax validation and  user intent matching. 
This multi-dimensional validation approach enables our benchmark to identify critical gaps in generated IaC template that remain undetected in syntax-only evaluations. 
As we report in Section~\ref{sec_evaluation}, the deployment validation reveals 42.7\% of the syntactically correct templates fail during deployment, highlighting the importance of testing beyond mere syntax. Similarly, the security validation identifies only 9.0\% of the deployable templates meet security compliance standards, exposing a critical shortcoming that is overlooked by IaC-Eval.


\subsection{Formative Study}



\begin{wrapfigure}{r}{0.53\linewidth} 
    \vspace{-15pt} 
    \setlength{\belowcaptionskip}{-10pt}
    \centering
    \includegraphics[width=\linewidth]{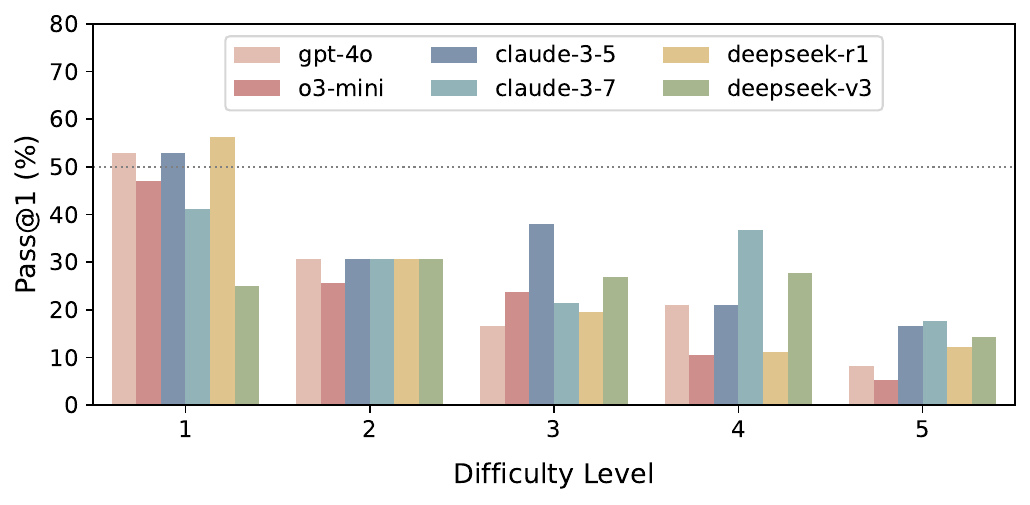}
    \captionsetup{justification=centering}
    \caption{Pass@1 scores across difficulty levels.}
    \label{fig:qq}
\end{wrapfigure}



We first conduct a simple evaluation of mainstream LLMs on IaC template generation task using DPIaC-Eval benchmark.
Fig.~\ref{fig:qq} presents the pass@1 accuracy results for six prominent LLMs across five difficulty levels. 
The results reveal that overall performance has significant room for improvement, with the best-performing models achieving only sightly above 50\% deployment success rates in level 1 difficulty.
Notably, all models exhibit a consistent inverse relationship between task complexity and performance, with score declining as difficulty increases from Level 1 to Level 5.
The average pass@1 at difficulty level 5 is just 12.8\%.
This pattern indicates that current LLMs struggle with the complexity of real-world infrastructure provisioning scenarios.
Thus, the preliminary results motivate us to propose a framework to fulfill this imperative research gap. 

\section{Framework}\label{sec_framework} 

LLMs have achieved remarkable success in code generation tasks~\cite{lachaux2020unsupervised, roziere2021leveraging, si2024solution, pan2024large}, demonstrating state-of-the-art performance across programming languages. However, their application to IaC generation remains limited and problematic. Current approaches~\cite{kon2024iac, ragothaman2024optimizing, pujar2023automated} treat IaC templates as conventional programs, focusing primarily on syntactic correctness while neglecting their continuous integration and deployment (CI/CD) nature~\cite{srivatsa2024survey, iac-ci-cd, liu2022first}. Consequently, generated templates often fail during actual deployment despite being syntactically valid and resulting in significantly lower accuracy rates compared to general programming tasks (GPT 4 achieves 19\% in IaC-Eval and 83\% in HumanEval). Our framework, \tool{}, addresses this limitation through a deployment-driven approach that enables LLMs to learn from deployment failures and continuously refine templates until achieving successful deployment. 
As shown in Fig.~\ref{fig:framework}, \tool{} consists of two modules, Iterative IaC Generation and IaC Deployment, integrated through a tiered feedback mechanism that mirrors real DevOps workflows. 
The complete prompts and other reproducible details mentioned in this section are available in our code repository.




\subsection{Iterative IaC Generation}\label{sec:II-A}
\begin{figure*}[t]
        \setlength{\abovecaptionskip}{5pt}
        \setlength{\belowcaptionskip}{-10pt}
        \centering
        \includegraphics[width=0.95\textwidth]{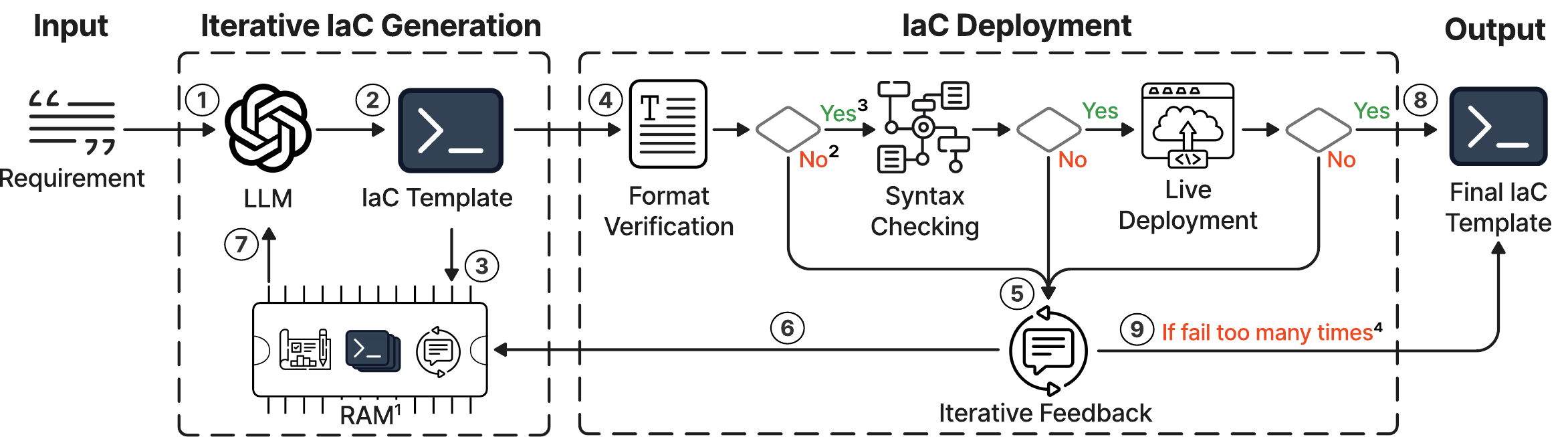}
        \caption{Workflow of \tool{}.$~^{1}$RAM: implemented as a local variable named \textit{conversation\_history} in Python code. 
        $~^{2}$No: indicates validation failure.$~^{3}$Yes: shows validation success.$~^{4}$Fail too many times: represents exceeding the maximum iteration threshold. 
        }
        \label{fig:framework}
\end{figure*}

Briefly, Iterative IaC Generation takes an IaC template description, written in natural language, as input and produces IaC template as output through a conversational feedback loop. The process works as follows and the numbers are indicated in Fig.~\ref{fig:framework}: 
\circled{1} A natural language prompt about the infrastructure requirement is fed to the LLM. 
\circled{2} The LLM generates an initial IaC template based on the requirement. 
\circled{3} The generated template and original prompt are stored in memory (implemented as a variable named \textit{conversation\_history} in code). 
\circled{4} This template is passed to the IaC Deployment module for multi-stage validation. 
\circled{5} If errors occur in IaC Deployment module, adaptive feedback is generated based on the specific failure point. We will elaborate on the adaptive feedback in Section~\ref{sec_framework_feedback}.
\circled{6} This feedback is stored in the conversation history along with previous interactions. 
\circled{7} The LLM is provided with both the feedback and all previous conversation history to generate an updated IaC template. 
To ensure output quality and consistency, in step 1, we apply a system prompt containing role assignment and task description, along with a lightweight chain-of-thought strategy to guide the generation process.
Steps 2 to 7 are repeated iteratively until either the IaC template successfully passes all three validation stages or the LLM exceeds the predetermined maximum number of allowed attempts. 

\subsection{IaC Deployment}\label{sec:II-B}
This module takes the LLM-generated IaC template as input and validates its soundness through a three-stage deployment.
Similarly to real-world testing, this process consists of sequential validations, from format verification to syntax checking to live deployment, with LLM iteratively refining the template based on error messages~\cite{iac-best-practices, vessey1985expertise, debugging-web}.
If an error occurs at any stage, feedback is generated and returned to the Iterative IaC Generation module. 
If the template passes all three validation stages, the framework will end with the current IaC template as the final output (step \circled{8}).

\paratitle{Format Verification}~Most IaC solutions use declarative definition files to specify infrastructure configurations. The syntax for describing IaC usually depends on the requirements of the target platform, with common formats including YAML and JSON. 
The first validation stage verifies whether the LLM-generated IaC template conforms to the correct format specification. 
As AWS CloudFormation templates written in YAML, we employ yamllint~\cite{yaml-linter} to validate basic structural correctness, including proper indentation, key-value pair formatting, and quotation usage.

\paratitle{Syntax Checking}~After confirming proper formatting, the second validation stage uses a static analysis tool, cfn-linter~\cite{cfn-linter}, to verify if the code is correct in CloudFormation syntax. 
Although most existing LLM-based IaC frameworks stop after this stage, our approach extends validation to actual deployment testing, which addresses a critical gap in operational viability assessment.

\paratitle{Live Deployment}~IaC shares a critical characteristic with traditional programming: code that passes compilation or syntax validation does not necessarily execute successfully at runtime. However, this challenge is particularly acute with IaC because of its unique properties.
Most IaC languages are declarative rather than imperative, specifying the desired end state rather than procedural steps. This fundamental difference creates a validation gap that static analysis alone cannot bridge~\cite{guerriero2019adoption}. Cloud environments impose numerous constraints that syntax validators cannot fully capture: regional service availability variations, deprecated resource identifiers, and service interdependencies. For example, a syntactically perfect Terraform template referencing an outdated Amazon Machine Image ID will pass linting but fail during actual provisioning. Thus, we employ live deployment validation with the AWS SDK boto3 client, the official toolkit developed by AWS.

 

\subsection{The Iterative Feedback}\label{sec_framework_feedback}
Previous research have shown the effectiveness of feedback mechanisms in code generation tasks~\cite{chen2023teaching, yang2022generating, huang2022large, zhou2024bridging}.
IaC deployment inherently involves continuous integration and development with cycles of validation and refinement~\cite{sokolowski2022infrastructure, aws-best-practices}.
By integrating a feedback loop mechanism into our framework, we simulate this real-world IaC development workflow while treating LLMs as ``human developers'', evaluating not only an LLM's initial generation capability, but also its error-resolution capacity.
Our feedback design implements a progressive assistance approach that mirrors established DevOps practices for IaC development. 
This iterative feedback reflects how cloud engineers typically engage with increasingly specific guidance when resolving deployment errors.

\paratitle{General Feedback} 
It provides high-level error indication by pointing out which validation stage failed during the multi-stage validation process described above.
Feedback follows a standardized structure:``\textit{Based on the evaluation, the template contains [type of error] Errors.}'' The types of error are YAML Syntax, CloudFormation Template Syntax, and Deployment. 
For example, when errors occur during the ``\textit{Formate Verification}'' stage, the LLM receives: ``\textit{Based on the evaluation, the template contains YAML Syntax Errors.}''. 
This mirrors the basic error alerts that developers encounter in CI/CD pipelines, requiring the LLM to diagnose and resolve issues with minimal guidance.

\paratitle{Detailed Feedback} It offers more specific guidance by including complete error messages returned by the validation tools (yamllint, cfn-lint, and AWS SDK). For instance, instead of simply indicating a deployment error, the LLM receives the full error message: ``\textit{An error occurred (ValidationError) when calling the CreateStack operation: Parameters: [KeyName] must have values.}'' This level of feedback approximates the detailed diagnostic information that developers typically encounter when working with IaC testing tools~\cite{hasan2020testing} and tests the LLM's ability to interpret technical error information.
Considering the experimental efficiency and practicability, we constrain feedback loops to a maximum of six iterations per stage, with two attempts for general feedback and four attempts for detailed feedback.
The asymmetrical allocation strategy is consistent with the empirical observations and experimental design in previous work~\cite{palavalli2024using}.

\section{Experimental Settings}\label{sec_exp_settings}
To comprehensively evaluate the performance of \tool{}, we conduct a thorough multidimensional evaluation guided by the following research questions:

\begin{itemize} [leftmargin=*, topsep=0pt] 
    \item \textbf{RQ1} \textit{How effective is \tool{} in IaC template generation?}
    \item \textbf{RQ2} \textit{What are the primary barriers of generating deployable IaC templates?}
    \item \textbf{RQ3} \textit{How effective is human-in-the-loop feedback strategy for \tool{}?}
    \item \textbf{RQ4} \textit{How trustworthy are the IaC templates generated by \tool{}?}
\end{itemize}

%
%

%
%

 
We select six representative leading LLMs, GPT-4o, GPT-o3-mini, Claude-3.5, Claude-3.7, DeepSeek-R1, and DeepSeek-V3.
Those models cover both open-sourced and closed-sourced ecosystems, broadly demonstrating competitive performance in general code generation tasks~\cite{chen2021evaluating, sobo2025evaluating, islam2025codesim, guo2024deepseek, tao2025privacy, gong2025towards} and have significant industry adoption~\cite{pan2024textit, deepseek-share}.
All evaluations are conducted through the APIs from official providers ensuring reproducibility and standardization. 
All models are evaluated with consistent temperature settings of 0 to maximize deterministic outputs and configured with the 8,000 maximum output token limit to accommodate complex IaC templates. 
We run those models on the benchmark introduced in Section~\ref{sec_benchmark}.

To ensure unbiased deployment evaluation, we create an isolated AWS account with default service quota and no pre-existing resources to mimic real production environment. 
Each validation cycle follows: allocating resources, validating their operational state, and then structurally removing all resources to set the environment back to its initial state. 
This approach ensures that each test runs in a clean environment, preventing ``cross-contamination'' between evaluation runs.

\paratitle{Evaluation Metrics}~The unbiased version of pass@k~\cite{chen2021evaluating} is a standard metric used by current benchmarks~\cite{chen2021evaluating, austin2021program, liu2023your}, which indicates the probability that at least one of the $k$ generated samples yields a correct solution.
However, it does not effectively capture the performance of iterative feedback. 
To address this limitation, we introduce passItr@n, where $n$ represents the maximum number of iterations allowed.
PassItr@n indicates the success rate that the IaC template is correctly generated (able to be deployed successfully) before or at iteration $n$.  






Unlike pass@k, which evaluates an LLM's ability to solve a problem by generating multiple independent solutions without feedback on correctness, passItr@n specifically measures how iterative feedback affects an LLM's ability to converge on correct solutions over time. 

\paratitle{Experiment costs}   
The cost of the study consists of two parts, one for the LLM API Token and one for the Deployment validations on AWS. The total cost (USD) is \$230.75 and \$35.21, respectively. 
On average, the token cost of the most expensive model (Claude-3.7-Sonnet) is \$0.42, but the cheapest model (DeepSeek-V3) is \$0.04 only, and the deployment cost is \$0.04, for each deployable template. These modest costs demonstrate that IaCGen provides a cost-effective approach for practical implementation with the lowest of \$0.08 to generate one deployable template. The AWS CloudFormation requires no fee for the deployment test. However, the deployment of some resources require fees once it is deployed such as the Relational Database Service.

\vspace{-6pt}

\section{Evaluation}\label{sec_evaluation}
\newcommand{\uparrowmark}{\textsuperscript{\textuparrow}}
\newcommand{\noarrowmark}{\textsuperscript{\phantom{$\textuparrow$}}}

\begin{figure}[t]
    \centering
    \begin{minipage}[]{0.48\textwidth}
        \renewcommand{\arraystretch}{1.3}
        \captionof{table}{Model performance with iterative feedback. 
        A column with an upward arrow\textsuperscript{↑} indicates its score is higher than the average. \textit{pItr} stands for \textit{passItr}.}
        \vspace{10pt}
        \label{tab:iteration}
        \resizebox{\linewidth}{!}{%
        \small
        \setlength{\tabcolsep}{6pt}
        \begin{tabular}{lcccc}
        \hline
        \rowcolor[HTML]{DADADA} 
        \textbf{Model} & \textbf{pItr@1} & \textbf{pItr@5} & \textbf{pItr@10} & \textbf{pItr@15} \\
        \hline
        \rowcolor[HTML]{F4F4F4} 
        \rowcolor[HTML]{F4F4F4} 
        GPT-4o          & 22.7\noarrowmark & 42.9\noarrowmark & 54.6\noarrowmark  & 55.2\noarrowmark \\
        \rowcolor[HTML]{DADADA} 
        GPT-o3-mini     & 20.8\noarrowmark & 39.0\noarrowmark & 66.2\noarrowmark  & 72.1\uparrowmark \\
        \rowcolor[HTML]{F4F4F4} 
        Claude-3.5      & \textbf{30.2}\uparrowmark & 64.9\uparrowmark  & \textbf{91.6}\uparrowmark  & \textbf{95.5}\uparrowmark \\
        \rowcolor[HTML]{DADADA} 
        Claude-3.7      & 26.8\uparrowmark & \textbf{66.7}\uparrowmark & 86.9\uparrowmark  & 92.8\textsuperscript{↑} \\
        \rowcolor[HTML]{F4F4F4} 
        DeepSeek-R1     & 22.9\noarrowmark & 53.6\uparrowmark  & 68.0\noarrowmark  & 71.9\noarrowmark \\
        \rowcolor[HTML]{DADADA} 
        DeepSeek-V3     & 24.2\noarrowmark & 42.5\noarrowmark  & 56.9\noarrowmark  & 61.4\noarrowmark \\
        \hline
        \rowcolor[HTML]{F4F4F4} 
        \textbf{Average} & 24.6\noarrowmark & 51.6\noarrowmark & 70.7\noarrowmark & 74.8\noarrowmark \\
        \hline
        \end{tabular}
        }
    \end{minipage}%
    \hfill
    \begin{minipage}[]{0.495\textwidth}
        \includegraphics[width=\linewidth]{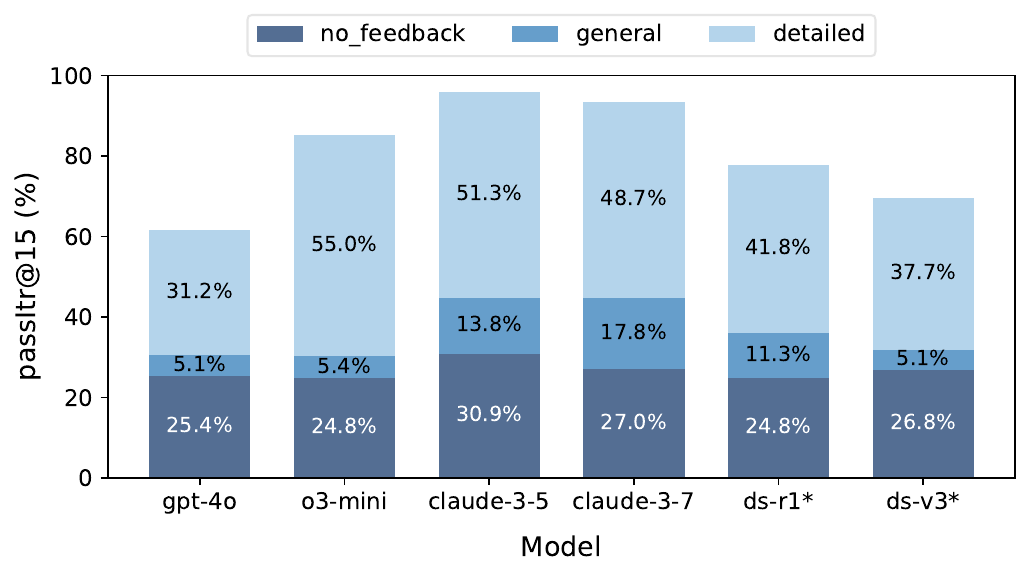}
        \caption{Distribution of maximum feedback level required for successfully generated templates. \textit{ds} stands for deepseek.}
        \label{fig:feedback_level_distribution}
    \end{minipage}
    \vspace{-10pt}
\end{figure}


\vspace{-1pt}
\subsection{RQ1: How effective is \tool{} in IaC template generation?}
Table~\ref{tab:iteration} presents the performance of our framework on the DPIaC-Eval benchmark as the iteration increases. 
The initial low success rates (passItr@1) align with the previous finding~\cite{kon2024iac}, with an average score of approximately 24.7\% on the first attempt of all models. 
This demonstrates the inherent challenge of LLM-based IaC template generation compared to general-purpose code generation tasks, where the average score of the models is 95.6\% in HumanEval benchmark~\cite{chen2021evaluating}.

Furthermore, all models achieve better performance with more iterations.
With sufficient iterations, all models eventually achieve a score higher than 50\%, and average of 75.2\% pass@Itr15 accuracy, which is a near 200\% performance improvement as compared to passItr@1. 
This shows \tool{} exhibits capability on effectively utilizing feedback to correct errors in IaC templates. 
Claude models show impressive performance in the generation of IaC templates, achieving leading results of over 90\% passItr@15 accuracy among the models assessed. 



\paratitle{Iterative Feedback}
Fig.~\ref{fig:feedback_level_distribution} shows the improvements with the iterative feedback mechanism for all models.
We observe that all models moderately benefit from general feedback (4.6\% to 17.6\%) and largely improve from detailed feedback (28.1\% to 51.0\%), highlighting that models can effectively learn from the iterative feedback. 
Claude models demonstrate remarkable performance, achieving a high passItr@15 score (95.5\% for Claude-3.5 and 92.8\% for Claude-3.7) 
In contrast, GPT-4o shows substantially lower performance (55.2\%).

\begin{figure}[t]
    \centering
    \begin{minipage}[t]{0.48\linewidth}
        \setlength{\abovecaptionskip}{-2pt}
        \setlength{\belowcaptionskip}{-13pt}
        \centering
        \includegraphics[width=\linewidth]{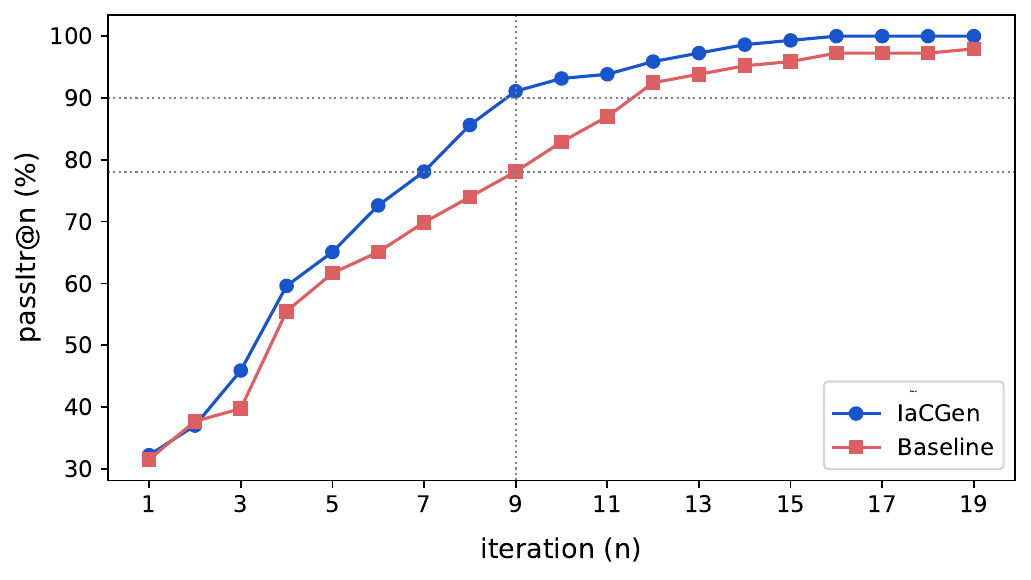}
        \caption{PassItr@n scores of \tool{} with and  without conversation history on Claude-3.5.}
        \label{fig:ablation_study}
    \end{minipage}
    \hfill
    \begin{minipage}[t]{0.48\linewidth}
        \setlength{\abovecaptionskip}{-2pt}
        \setlength{\belowcaptionskip}{-12pt}
        \centering
        \includegraphics[width=\linewidth]{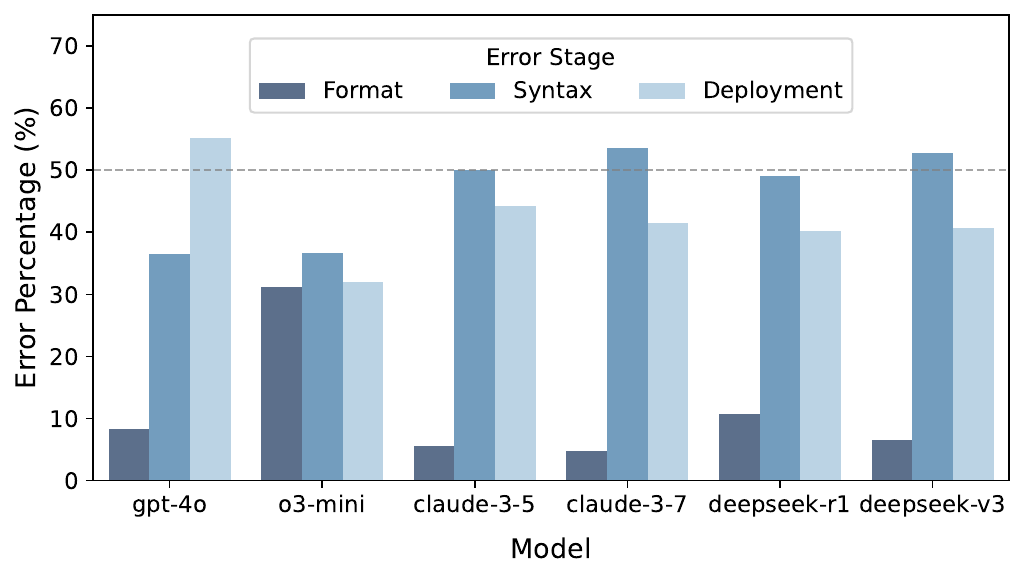}
        \caption{Distribution of error stages (\textit{i.e.,} where the generated template fails) across models.}
        \label{fig:er}
    \end{minipage}
    \vspace{8pt}
\end{figure}

\label{content:ablation_study}
\paratitle{Effectiveness of Conversation History} 
Unlike previous research~\cite{kon2024iac, palavalli2024using, ragothaman2024optimizing, madaan2023self} which isolates each feedback iteration, our approach preserves the entire conversation history for each iteration. 
To evaluate the effectiveness of our iterative feedback mechanism that provides LLMs with the entire conversation history, we conduct an ablation study that compares it to the baseline approach, which only feeds the system prompt, original prompt, error message, and incorrect IaC template to the LLM. 
We select Claude-3.5, the best performing model, for this study.
As illustrated in Fig.~\ref{fig:ablation_study}, both approaches demonstrate comparable performance during the first two iterations. 
However, our approach establishes a consistent advantage starting with the third iteration. 
Our method achieves a deployable IaC template in an average of 4.55 iterations, while the baseline approach required an average of 5.41 iterations, which is an 15.9\% reduction in required iterations.

Without accessing the entire conversation history, we find that LLMs frequently reintroduce previously corrected errors when addressing new issues. 
This ``Error Recurrence'' phenomenon suggests that maintaining the conversation history provides a crucial context for LLMs to learn from past corrections and avoid repeating similar mistakes across iterations.
This improvement might stem from IaC's inherent complexity, which deployment errors often result from intricate resource interdependencies rather than isolated issues, yet error messages typically identify only the immediate failure point. 
By providing complete conversation history, LLMs can understand the cascading effects of their previous modifications and make more informed corrections. 
This contextual awareness enables systematic problem-solving rather than random attempts, as evidenced by the reduced error recurrence and faster convergence to deployable templates. 

\paratitle{Generalizability of \tool{}}
IaCGen's modular design and the cross-language capabilities of LLMs enables seamless adaptation to different IaC technologies. Practitioners can adapt IaCGen to alternative IaC languages and cloud service providers by substituting the corresponding linters (for format and syntax validation) and deployment components (for deployment verification).
To evaluate \tool{}'s generalizability quantitatively, we conduct experiments on the IaC-Eval benchmark~\cite{kon2024iac}, which contains Terraform templates, with updates to the linter. 
Due to fundamental architectural differences between CloudFormation and Terraform (discussed in Section~\ref{sec:difference-between-IaC-tools}), we focus on syntax validation following the established methodology of IaC-Eval, using Claude-3.5 based on its superior performance in our primary experiments. \tool{} demonstrates strong generalizability to Terraform, achieving 79.7\% passItr@1 and 100\% passItr@7 accuracy with an average of 1.58 iterations. For comparison, \tool{} requires 2.53 iterations on average to achieve 100\% passItr@5 accuracy on DPIaC-Eval's CloudFormation templates. This difference reflects the higher complexity of DPIaC-Eval templates, evidenced by their higher average lines of code and resource complexity.



\vspace{-2pt}

\find{{\bf Answer to RQ1:}~\tool{} demonstrates high effectiveness in IaC template generation. Through iterative feedback, it transforms the initial poor success rates ($\approx$24.7\%) into high deployment capabilities ($\approx$75.2\%) for all evaluated models, and Claude models achieve over 90\% passItr@15. IaCGen also shows its strong generalizability to Terraform with $\approx$25\% performance improvement. Maintaining complete conversation history can effectively mitigate the ``Error Recurrence'' phenomenon.}

\subsection{RQ2: What are the primary barriers of generating deployable IaC templates?}
Fig.~\ref{fig:er} shows the distribution of error stage where LLMs encounter the first deployability error during IaC generation. 
Syntax issues, which occur during the syntax checking stage, constitute the primary challenge for nearly all models, followed by deployment errors that only manifest during live deployment testing. This finding validates our methodological focus on the validation of deployability beyond mere syntactic correctness, highlighting the gap between code that appears valid and code that functions correctly in production environments.

OpenAI's models exhibit distinct error patterns compared to other models in our evaluation. While GPT-4o's primary challenge is deployment errors (suggesting strong syntax capabilities but weaker understanding of operational constraints), GPT-o3-mini experiences a disproportionate rate of formatting errors ($>$30\% compared to approximately 10\% for other models). This discrepancy points to fundamental differences in how various model architectures approach structured template generation, potentially reflecting differences in training data composition or model architecture.

We identify five primary error categories with notable frequency and impact on model performance.
Table~\ref{tab:error_message} illustrates the error messages and the distribution among models. 

\newcommand{\figWidth}{0.05\textwidth}
\newcommand{\figHeight}{5mm}

\setlength{\textfloatsep}{8pt}   

\begin{table*}[t]
\centering
\caption{Five most common errors across three validation stages. xx$^1$: placeholder for line number and value name. For \textit{Count}, from left to right are: GPT-4o, o3-mini, Claude-3.5, Claude-3.7, DeepSeek-R1, DeepSeek-V3.}
\label{tab:error_message}
\renewcommand{\arraystretch}{1.3}
\resizebox{0.99\textwidth}{!}{%
\begin{tabular}{
    >{\centering\arraybackslash}l
    |>{\centering\arraybackslash}c
    |>{\centering\arraybackslash}m{8.5cm}
    |>{\centering\arraybackslash}c
    |>{\centering\arraybackslash}c
    |>{\centering\arraybackslash}c
}
\hline

\rowcolor[HTML]{DADADA}
\textbf{\#} & \textbf{Error Type} & \textbf{Error Message} & \textbf{Percentage} & \textbf{Failed Stage} & \textbf{Count} \\ 
\hline

\rowcolor[HTML]{F4F4F4} 
1 & Missing Value & An error occurred (ValidationError) when calling the CreateStack operation: Parameters: {[}xx$^1${]} must have values & 14.69\% & Deployment & \raisebox{-0.5\height}{\includegraphics[width=\figWidth, height=\figHeight]{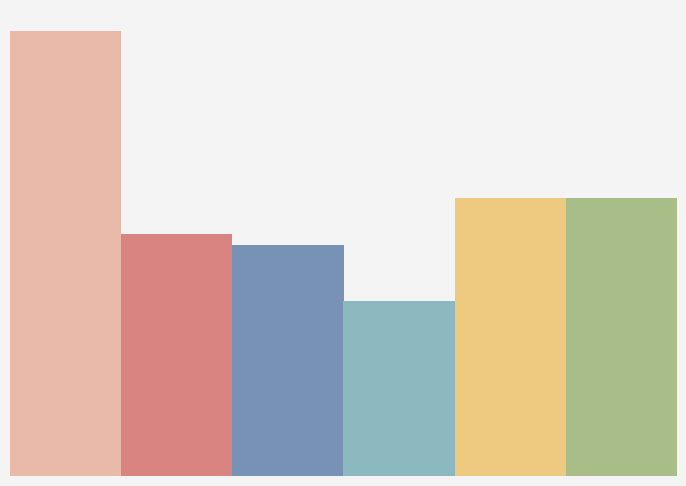}} \\

\rowcolor[HTML]{DADADA} 

2 & Self-defined Property & Additional properties are not allowed (`xx' was unexpected) & 6.44\% & Syntax &  \raisebox{-0.5\height}{\includegraphics[width=\figWidth, height=\figHeight]{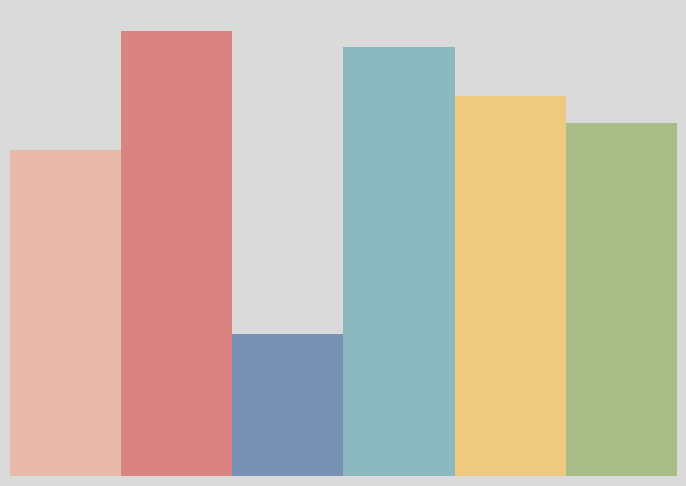}} \\

\rowcolor[HTML]{F4F4F4} 
3 & Null Substitution & `Fn::Sub' isn't needed because there are no variables & 3.09\% & Syntax &  \raisebox{-0.5\height}{\includegraphics[width=\figWidth, height=\figHeight]{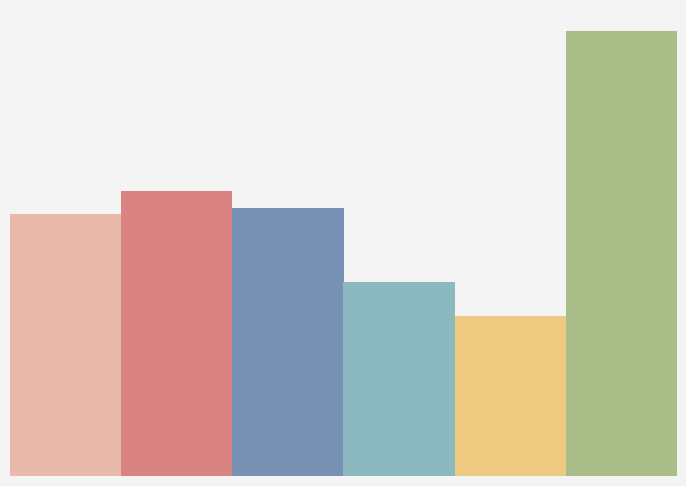}} \\

\rowcolor[HTML]{DADADA} 
4 & Unnecessary Whitespace & Line xx: too many spaces inside brackets & 2.92\% & Format &  \raisebox{-0.5\height}{\includegraphics[width=\figWidth, height=\figHeight]{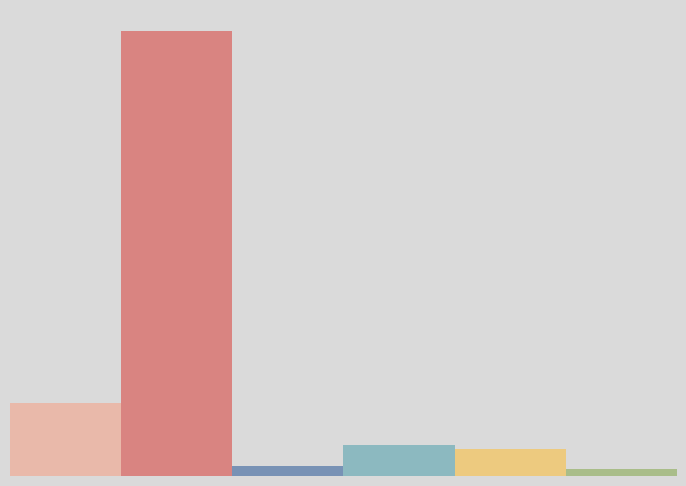}} \\

\rowcolor[HTML]{F4F4F4} 
5 &  Arbitrary Default Value & Parameter validation failed: parameter value xx for parameter name xx does not exist & 1.19\% & Deployment &  \raisebox{-0.5\height}{\includegraphics[width=\figWidth, height=\figHeight]{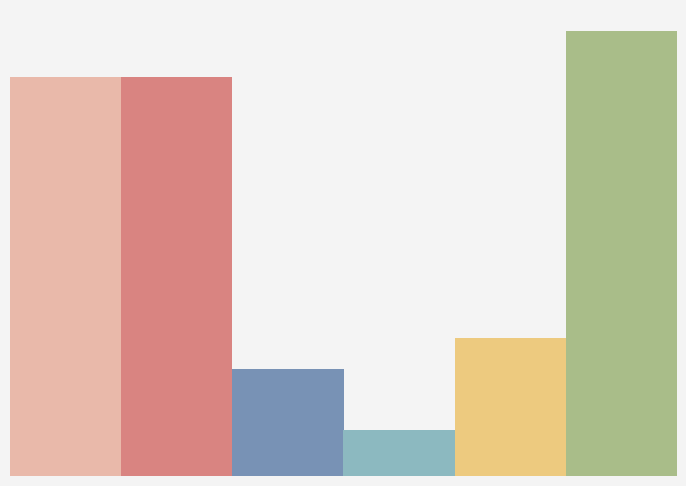}} \\

\hline
\end{tabular}}
\end{table*}

\textbf{(1) Missing Value.}
The most prevalent (14.70\%) error across models is the deployment error: ``\textit{An error occurred (ValidationError) when calling the CreateStack operation: Parameters: [xx] must have values.}'' 
It happens when LLMs define parameters without providing corresponding values, which is particularly common with the \textit{KeyName}, \textit{VPC}, and \textit{Subnet} parameters.
Sometimes, models deliberately leave parameters undefined when information is not explicitly provided in the prompt (e.g., email addresses for SNS notifications). 
This error also appears to be models' assumption that resources have already existed in the cloud environment, leading them to create placeholders for users to manually fulfill. 
Among all models, Claude-3.7 demonstrates the lowest occurrence rate of this error and the highest success rate in resolving it through iterative feedback.
This pattern could stem from the prevalence of such template structures in the GitHub repositories as we observe. 

\textbf{(2) Self-defined Property.}
The second most frequent (6.44\%) error manifests itself as: ``\textit{Additional properties are not allowed (`xx' was unexpected).}'' 
This syntax error occurs when models hallucinate non-existent attributes for AWS services or incorrectly use properties from other AWS services. Claude-3.5 shows remarkable resilience to this type of error, with only 37 occurrences compared to an average of 101 occurrences in other models.

\textbf{(3) Null Substitution.}
``\textit{`Fn::Sub' is not needed because there are no variables''} is the third most common (3.09\%) error. This error typically occurs in the \textit{UserData} attributes of ``\textit{AWS::EC2::Instance}'' and ``\textit{AWS::EC2::LaunchTemplate}'', where the models employ \textit{``!Sub $|$''} despite the absence of variables that require substitution. 
This pattern likely results from large-scale training corpus that contain numerous examples using this function with \textit{UserData}, leading models to adopt it as a standard practice regardless of necessity.

\textbf{(4) Unnecessary Whitespace.}
``\textit{Line xx: too many spaces inside brackets''} accounts for 2.92\% of all errors. 
This issue disproportionately affects o3-mini, accounting for 75\% of all instances across models.
The error typically manifests itself in constructs like ``\textit{!Select [ 0, !GetAZs `' ]}'' (Note the whitespaces on the left of `\textit{0}' and `\textit{]}') instead of the correct way of ``\textit{!Select [0, !GetAZs `']}''. o3-mini consistently inserts extraneous whitespace adjacent to brackets. This error is particularly resistant to correction, often requiring explicit prompt rather than general instructions to resolve.

\textbf{(5) Arbitrary Default Value.}
``\textit{Parameter validation failed: parameter value xx for parameter name xx does not exist''} denotes the error that occurs when models attempt to resolve missing parameter values by providing arbitrary defaults. 
Claude-3.5, Claude-3.7, and DeepSeek-R1 demonstrate superior resistance in this category, averaging only 6 occurrences per model compared to approximately 27 occurrences for other models. 
This suggests that these models hold a more comprehensive understanding of the appropriate default values for the AWS resource parameters.

\vspace{-2pt}
\find{{\bf Answer to RQ2:}
Most of errors occur during syntax stage, slightly more than those identified during deployment errors.
The most common error type is \textit{Missing Value}, and the top five error types account for around 30\% of the total failure instances. 
Different models exhibit a distinct error pattern, and Claude models show the best performance in handling most of the common errors.
}

%
%

%
%
\subsection{RQ3: How effective is human-in-the-loop feedback strategy for \tool{}?} 
Human-in-the-loop feedback has emerged as a crucial strategy across machine learning and LLM domains~\cite{mosqueira2023human, zanzotto2019human, xiao2023llm}, demonstrating significant improvements in complex scenarios especially in SE tasks. 
Enlightened by this, we incorporate human feedback into our framework to explore its capability on tackling unique challenges of IaC template generation. This level of feedback simulates the guidance that a senior cloud engineer might provide to a junior developer during code review, testing whether LLMs can effectively utilize and incorporate direct expert instructions.

To comprehensively evaluate the effectiveness, we include human feedback in addition to the general and detailed feedback mentioned in Section~\ref{sec_framework_feedback}. 
Human feedback represents the highest level of assistance in the three-tier feedback hierarchy, providing explicit correction guidance such as: \textit{``Add CidrBlock: 10.0.0.0/16 to the VPC resource''} or \textit{``Replace the AMI ID with {placeholder for correct AMI ID value}.''} 
A cloud engineer is asked to craft the human feedback based on the 
reference 
template and error messages. 
We assign the same four attempts for the human feedback as the detailed feedback with similar rationale.
Therefore, the maximum attempts for each stage (Format Verification, Syntax Checking, and Live Deployment) will be ten iterations.


\begin{wrapfigure}{r}{0.53\linewidth} 
    \vspace{-10pt} 
    \setlength{\abovecaptionskip}{-2pt}
    \setlength{\belowcaptionskip}{-2pt}
    \centering
    \includegraphics[width=\linewidth]{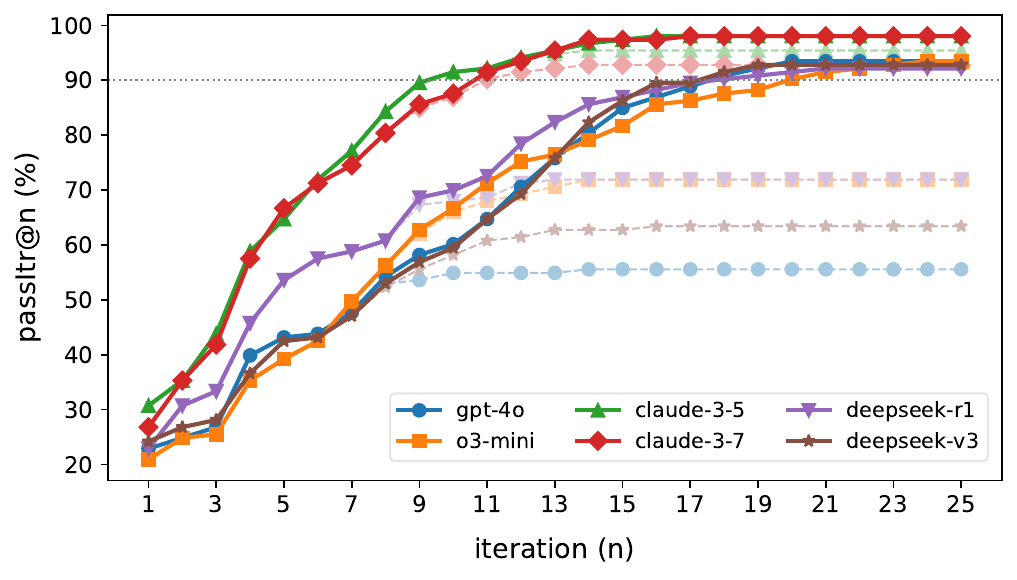}
    \caption{PassItr@n scores for evaluated models. Transparent lines refer to the result with general and detailed feedback. Solid lines refer to the results additionally with human feedback.}
    \label{fig:all_model_linear_graph}
\end{wrapfigure}

With human feedback, all evaluated models achieved a passItr@25 accuracy higher than 90\% and the Claude models achieve 98.0\% accuracy as shown in Fig.~\ref{fig:all_model_linear_graph}. The results demonstrate that human intervention provides consistent improvements across different model architectures. 
Moreover, our analysis reveals that human feedback is particularly effective in resolving error categories that LLMs struggle to address independently. 
Notably, human intervention can successfully resolve persistent issues such as the \textit{``Missing Value''} errors (as identified in RQ2). This phenomenon is especially significant for GPT-o3-mini, which demonstrates considerable difficulty in resolving ``\textit{Unnecessary Whitespace}'' errors without explicit human guidance. The model's tendency to insert extraneous whitespace proves to be resistant to automated feedback correction but responds effectively to direct human instruction specifying the correct formatting pattern. Similarly, human experts can provide domain-specific knowledge about appropriate parameter values and resource configurations for \textit{``Arbitrary Default Value''} errors, information that may not be apparent from automated error messages alone. This targeted intervention demonstrates the complementary nature of human expertise and automated systems in complicated SE tasks.

\find{{\bf Answer to RQ3:}
Human feedback provides consistent improvements across all evaluated models, with gains ranging from 2.6\% to 37.9\% percentage points. All models achieve a performance of over 90\% passItr@25, and the Claude models can achieve 98\% passItr@25.
}  

\subsection{RQ4: How trustworthy are the IaC templates generated by \tool{}?}\label{sec:IV-C}

In previous RQs, we examine the syntactic correctness and deployability of the generated IaC templates. 
To ensure their practical utility in real-world scenarios, we further assess the trustworthiness of the generated templates along two critical dimensions: User Intent Matching and Security. 

\paratitle{User Intent Matching}
User requirements for IaC templates are often ambiguously specified due to the inherent complexity and technical nature of cloud infrastructure.
To evaluate how well the generated IaC templates align with the intended infrastructure requirements, we employ an intent match evaluation. 
Specifically, based on the DPIaC-Eval benchmark, a cloud engineering expert formulates intent specifications using Checkov’s custom policy\ty{~\cite{checkov-custom-policy}}, comprising two elements:
\begin{itemize} [leftmargin = *, topsep=0pt]
    \item Required Resources: the core infrastructure components that must be presented in the template.
    \item Required Attributes: the essential configuration parameters associated with these resources.
\end{itemize}

To manage the annotation workload, we randomly sample 51 instances from the benchmark for this evaluation. The generated IaC templates are then systematically compared against the manually crafted intent specifications to assess user intent matching to the stated requirements.

We simply use the coverage (\%) to denote the level of user intent matching, \textit{i.e.,} how many required resources or required attributes are included in the generated templates. 
Table~\ref{tab:user_intent} shows the varying performance among models on capturing the intention of users.
We observe a relatively sound capability on generating the required resources, with an average of 58.8\% templates that include all specified resources across all models.
In contrast, attribute-level matching presents a greater challenge, with models achieving an average of only 40.5\% compliance in generating all required attributes with corresponding values. 
Only 25.2\% of the templates satisfy both the resource-level and attribute-level requirements across all models.
Claude-3.7 demonstrates superior overall performance, yet only 31.4\% of its generated IaC templates fully satisfy the user intent.

\paratitle{Security}
Security of IaC templates represents a critical concern in cloud computing, as insecure configurations can lead to data breaches, unauthorized access, and compliance violations~\cite{drosos2024your}. 
To evaluate the security posture of LLM-generated IaC templates that passed deployability evaluation, we perform the security analysis by employing Checkov~\cite{checkov}, an popular open source policy-as-code tool for IaC compliance scans~\cite{minna2025analyzing} with common industry standards such as the Center for Internet Security~\cite{cis} and AWS Foundations Benchmark~\cite{cis-aws}. We define three distinct compliance metrics:
\begin{itemize} [leftmargin = *, topsep=0pt]
    \item \textit{Policy Pass} Rate (\%): The percentage of security policies satisfied across all security policies checked by Checkov.
    \item \textit{Unfiltered Compliance} Rate (\%): The proportion of generated templates that pass all applicable Checkov security policy checks, irrespective of policy applicability.
    \item \textit{Filtered Compliance} Rate (\%): To account for Checkov's limited policy coverage, this metric excludes templates to which no relevant security policies can be applied. 
\end{itemize}

%
%
\newcolumntype{P}[1]{>{\centering\arraybackslash}p{#1}}

\begin{table*}[t]

\centering

\begin{minipage}[b]{0.48\linewidth}
    \centering
    \caption{User intent performance (\%) of the generated templates at maximum iterations across models. Resource stands for resource level intent match and Attribute stands for attribute level intent match.}
    \label{tab:user_intent}
    \resizebox{\linewidth}{!}{%
    \begin{tabular}{p{2cm} P{2cm} P{2cm} P{2cm}}
    \hline
    \rowcolor[HTML]{DADADA} 
    \textbf{Model} 
        & \textbf{\makecell{Resource}} 
        & \textbf{\makecell{Attribute}} 
        & \textbf{\makecell{Resource \&\\Attribute}} \\
    \hline
    \rowcolor[HTML]{F4F4F4} 
    GPT-4o         & 51.0\noarrowmark & 41.2\textsuperscript{↑} & 21.6\noarrowmark \\
    \rowcolor[HTML]{DADADA} 
    GPT-o3-mini    & 58.8\noarrowmark & 35.3\noarrowmark & 23.5\noarrowmark \\
    \rowcolor[HTML]{F4F4F4} 
    Claude-3.5     & 60.8\textsuperscript{↑} & \textbf{47.1}\textsuperscript{↑} & 25.5\textsuperscript{↑} \\
    \rowcolor[HTML]{DADADA} 
    Claude-3.7     & 60.8\textsuperscript{↑} & 39.2\noarrowmark & \textbf{31.4}\textsuperscript{↑} \\
    \rowcolor[HTML]{F4F4F4} 
    DeepSeek-R1    & \textbf{64.7}\textsuperscript{↑} & 39.2\noarrowmark & 23.5\noarrowmark \\
    \rowcolor[HTML]{DADADA} 
    DeepSeek-V3    & 56.9\noarrowmark & 41.2\textsuperscript{↑} & 25.5\textsuperscript{↑} \\
    \hline
    \rowcolor[HTML]{F4F4F4} 
    \textbf{Average} & 58.8\noarrowmark & 40.5\noarrowmark & 25.2\noarrowmark \\
    \hline
    \end{tabular}
    }
\end{minipage}
\hfill
\begin{minipage}[b]{0.48\linewidth}
    \centering
    \caption{Security performance (\%) of the generated templates at maximum iterations across models.  Higher numerical values indicate stronger security. \textit{Compl.} stands or \textit{Compliance}.}
    \label{tab:security}
    \resizebox{\linewidth}{!}{%
    \begin{tabular}{p{2cm} P{2cm} P{2cm} P{2cm}}
    \hline
    \rowcolor[HTML]{DADADA} 
    \textbf{Model} 
        & \textbf{\makecell{Policy\\Pass}} 
        & \textbf{\makecell{Unfiltered\\Compl.}} 
        & \textbf{\makecell{Filtered\\Compl.}} \\
    \hline
    \rowcolor[HTML]{F4F4F4} 
    GPT-4o         & 71.0\noarrowmark & 19.0\noarrowmark & 6.1\noarrowmark \\
    \rowcolor[HTML]{DADADA} 
    GPT-o3-mini    & 71.9\noarrowmark & 18.3\noarrowmark & 7.4\noarrowmark \\
    \rowcolor[HTML]{F4F4F4} 
    Claude-3.5     & 78.0\textsuperscript{↑} & 17.6\noarrowmark & 6.7\noarrowmark \\
    \rowcolor[HTML]{DADADA} 
    Claude-3.7     & \textbf{78.9}\textsuperscript{↑} & 21.6\textsuperscript{↑} & 10.4\textsuperscript{↑} \\
    \rowcolor[HTML]{F4F4F4} 
    DeepSeek-R1    & 75.1\noarrowmark & 20.3\textsuperscript{↑} & 8.3\noarrowmark \\
    \rowcolor[HTML]{DADADA} 
    DeepSeek-V3    & 76.8\textsuperscript{↑} & \textbf{24.2}\textsuperscript{↑} & \textbf{11.5}\textsuperscript{↑} \\
    \hline
    \rowcolor[HTML]{F4F4F4} 
    \textbf{Average} & 75.3\noarrowmark & 20.2\noarrowmark & 8.4\noarrowmark \\
    \hline
    \end{tabular}
    }%
\end{minipage}
\end{table*}
%
%
Table~\ref{tab:security} shows the security compliance of generated IaC templates, and it reveals an alarmingly low security compliance rate overall.
In terms of security policies, 75.3\% of the examined security policies are passed on average. 
Surprisingly, at template-level, only 20.2\% unfiltered compliance and 8.4\% filtered compliance are obtained.
This apparent contradiction suggests that, while models generally adhere to most security rules, critical failures in at least one policy consistently invalidate entire templates.
DeepSeek-V3 demonstrates the best performance in both unfiltered and filtered compliance (24.2\% and 11.5\%). Claude-3.5 ranks the lowest (17.6\%) in unfiltered compliance and GPT-4o shows the weakest in filtered compliance (6.1\%).
Further analysis reveals 55.1\% of template failures originate from three AWS services: \textit{S3 Bucket}, \textit{Lambda Function}, and \textit{EC2 SecurityGroup.} These findings indicate a critical threshold effect. While models successfully implement most security requirements ($>$70.0\% policy pass), failures in 1-2 policies consistently invalidate entire templates. The discrepancy between high policy-level compliance and catastrophic template-level failure highlights fundamental challenges in the ability of LLMs to generate secure IaC templates.


\vspace{-2pt}

\find{{\bf Answer to RQ4:}
The trustworthiness of LLM-generated IaC templates remains moderate.
Only 25.2\% generated templates fully satisfy the user intent, with reasonable resource identification (58.8\%) but weaker attribute fulfillment (40.5\%). 
Security analysis shows that despite a 75.3\% policy pass rate, only 8.4\% of the templates achieved full compliance.
}




\section{Discussion}\label{sec_discuss}

\subsection{Security Policy Violations in IaC Templates}
As demonstrated in our RQ4 evaluation, current models exhibit limited sensitivity to security concerns, potentially leading to vulnerabilities such as data breaches and service outages.
Our security analysis identified 70 distinct policy violations across all generated IaC templates, with a severity distribution of 36 low-severity, 17 medium-severity, 7 high-severity, and 10 informational-level violations according to Prisma Cloud classifications~\cite{prisma_cloud}.
We further analyze the top 10 most frequently violated policies, revealing one medium-severity, seven low-severity, and two informational-level violations. 
This distribution indicates that LLMs not only fail to implement security best practices but also potentially introduce critical vulnerabilities.


The medium-severity policy: ``\textit{AWS SNS topic has SSE disabled}'', reflects one systematic security oversight where LLMs fail to enable server-side encryption for message queuing services~\cite{medium_severity_policy} in generated templates. 
This omission exposes sensitive data in transit and at rest, violates data protection principles, and potentially compromises confidential information processed through messaging systems. 
In addition, violation of the low-severity policy: ``\textit{AWS security groups allow ingress from 0.0.0.0/0 to port 80}'', can expose AWS resources to potential security threats~\cite{low_severity_policy}.
Notably, low-severity does not mean it necessarily causes low security impact~\cite{dahabiyeh2021ignorance, low_severity_impact}.
Although classified as low-severity, this configuration creates a direct attack vector by allowing traffic from all IP addresses to the web service port. Such misconfiguration can facilitate reconnaissance attacks, denial-of-service attempts, and unauthorized access to application endpoints~\cite{mathew2014study}.

The security policy violations could stem from the problematic training data. 
Analysis of publicly available IaC templates reveals that many contain insecure default configurations, such as security groups configured with \textit{``0.0.0.0/0''} IP ranges as placeholder values~\cite{public_problematic_template_security_group}, or SNS topics deployed without encryption keys for operational simplicity~\cite{public_problematic_template_SNS}.
LLMs appear to be pre-trained and perpetuate these problematic patterns without understanding their security implications.

For researchers, the security policy violations underscore the need to develop AI systems that integrate security considerations into their core design. 
Our work demonstrates the necessity for security-focused evaluation metrics in AI4SE research. Future benchmarks could incorporate multi-dimensional security assessments, including not only static policy compliance but also dynamic security analysis and threat modeling capabilities. 
Furthermore, practitioners should implement robust validation mechanisms before adopting LLM-generated IaC templates, to prevent security vulnerabilities from reaching production environments.

\subsection{Human Feedback: Automation Potential}
Even though the Claude models can achieve satisfying results (94.15\% passItr@15) without human intervention, results show human feedback can enhance the overall performance. 
To explore the possibility on automation human feedback, we conduct a post-hoc analysis.
We surprisingly find that most of human feedback could be derived from publicly available sources, such as AWS documentation~\cite{aws-doc}, Stack Overflow~\cite{stackoverflow} discussions, and technical blogs.
Three representative cases demonstrate how human feedback can be obtained from crowd knowledge:



\textit{S3 Bucket Naming Conflicts.} GPT models commonly assign generic bucket names, causing deployment failures due to S3's global uniqueness requirement.
The original human feedback is: ``\textit{S3 bucket names must be globally unique across all AWS accounts. Consider adding a unique identifier such as: random string, organization prefix, timestamp, or environment identifier to ensure uniqueness}'', which guides the model to generate unique names following established naming conventions. 
This principle directly corresponds to the best practices documented in technical resources~\cite{bucket_naming_solution, bucket_naming_guide}, which explicitly stipulate name uniqueness (\textit{``Bucket names must be unique across all AWS accounts in all of the AWS''}) and solutions (\textit{``Include a Randomized Suffix or Prefix''}).

\textit{Outdated AMI References.} LLMs frequently provide invalid AMI identifiers for EC2 instances probably caused by biased and outdated training data, as AMI changes and is different between regions. 
Human feedback directs models to refer to AWS Systems Manager Parameter Store for dynamic AMI resolution. This approach follows established instructions documented in official AWS blogs~\cite{ami_solution}, which demonstrate programmatic AMI lookup using the latest Amazon Linux AMI parameters (\textit{``Querying for the latest AMI using public parameters''}).

\textit{Service-Specific Configuration Constraints.} Complex AWS service interdependencies often result in attribute-level errors, such as the constraint ``\textit{DbSubnetGroupName should not be specified for read replicas created in the same region as the master.}'' Human feedback resolves this by explicitly instructing LLMs to remove the conflicting attribute. 
Similarly, the solution is directly related to the AWS documentation~\cite{dbinstance_solution}, which explicitly describes this constraint in the DbSubnetGroupName specification (\textit{``The specified DB subnet group must be in the same AWS Region...''}).

Our empirical observation suggests that future iterations of IaCGen could achieve human-in-the-loop performance through LLM-based knowledge retrieval mechanisms (e.g., agentic framework, Retrieval-Augmented Generation) while maintaining full automation. 

\vspace{-5pt}
\subsection{DevOps-Inspired LLM-based Framework}

Our findings suggest that iterative feedback-driven approaches grounded in common software development principles such as CI/CD, Agile, and Scrum yield superior outcomes in IaC template generation. Rather than employing heavy prompting engineering or relying on task-specific heuristics, our framework systematically incorporates iterative feedback to guide the model across multiple refinement stages. This iterative structure mirrors conventional SE practices, such as progressive refinement, requirement validation, and configuration testing. 

The alignment with these established methodologies not only improves model outputs, but also enhances interpretability and developer trust.
These DevOps-derived heuristics may serve as a robust and generalizable research paradigm for broader AI4SE tasks, bridging the gap between LLM capabilities and practical software engineering workflows. 
The success of this approach suggests that future AI4SE frameworks could focus on iterative refinement over single-shot generation for practical performance enhancement, particularly for complex, multi-constraint problems.

\vspace{-5pt}
\subsection{Threats to Validity}
While we believe this study makes substantial contributions to LLM for IaC  generation, several limitations should be acknowledged. 
First, our evaluation includes models available at the time of writing. 
The rapid evolution of LLMs may yield different performance patterns with newer releases. 
However, our framework's model-agnostic design ensures compatibility with future developments. 
To facilitate reproducibility, we document the specific model and their cut-off date in our replication package.
Second, our benchmark covers 153 scenarios across 58 commonly used AWS services, representing substantial coverage of typical infrastructure patterns. 
However, highly specialized configurations may not be fully captured. 
Also, our difficulty categorization, while systematic, may not align perfectly with all organizational perspectives on complexity.
Lastly, while a comprehensive evaluation across all IaC languages would be ideal, the absence of standardized benchmarks and the substantial effort required to create such domain-specific evaluation datasets limits the broader assessment. 
However, our demonstrated generalizability to Terraform and IaCGen's modular architecture enable straightforward extension to other IaC ecosystems. 
\section{Related Work}

\subsection{Feedback Mechanism of LLMs}
Feedback mechanism has been widely adopted to enhance LLM outputs in SE tasks. 
For instance, Chen et al.~\cite{chen2023teaching} used natural language feedback to refine code completions, Huang et al.~\cite{huang2022large} introduced feedback signals during model training, and Yang et al.~\cite{yang2022generating} leveraged feedback for iterative output correction. Despite these advances, the use of feedback in IaC generation remains underexplored, particularly when considering the need for executable and deployable outputs. 

Several studies have incorporated iterative feedback into LLM-based code generation~\cite{zhang2023repocoder, bi2024iterative, bouzenia2025you}. 
However, these approaches typically provide only immediate error messages and problematic code snippets. 
Due to IaC's declarative nature and complex resource interdependencies, addressing one error often introduces new issues elsewhere, as fixes frequently require coordinated changes across multiple template sections. 
Thus, in this study, we provide the complete conversation history to the LLM, including all previous errors and attempted solutions, enabling in-context learning~\cite{dong2022survey} and a more informed decision-making that avoids repeating previous mistakes.

\vspace{-5pt}
\subsection{LLMs for IaC Generation}
Despite the growing importance of IaC in modern cloud development, research on LLMs' capability to generate reliable infrastructure code remains limited. 
Kon et al.~\cite{kon2024iac} introduced the first benchmark for evaluating LLM-generated IaC, focusing primarily on syntactic validity and intent compliance across Terraform templates. 
They found LLMs achieved less than 20\% success rates, which is substantially lower than general code generation tasks. 
Ragothaman~\cite{ragothaman2024optimizing} attempted to address deployability concerns but provided limited technical details about their validation process and relied heavily on manual testing procedures. While they identified the potential value of incorporating error feedback from failed templates into prompts or retrieval-augmented generation systems, they did not implement or evaluate such mechanisms. Similarly, Palavalli et al.~\cite{palavalli2024using} examined feedback loops for IaC generation, discovering that feedback effectiveness diminishes significantly after five iterations. However, their evaluation remained limited to syntax validation, and their feedback strategy only provided incorrect IaC template and basic error information.

Compared to existing research, our work advances this research area through three key contributions. 
We introduce the first comprehensive deployability-centric benchmark that validates templates in live cloud environments, capturing the complete spectrum of real-world deployment failures. 
Additionally, our framework substantially improves performance through deployment-driven iterative refinement. 
Finally, by prioritizing deployability, our approach provides the first practical evaluation of LLM capabilities on IaC generation in DevOps.

\vspace{-8pt}
\section{Conclusion}
\vspace{-2pt}
IaC generation holds significant promise for automating cloud infrastructure provisioning in DevOps.
We construct DPIaC-Eval, a deployability-centric IaC template benchmark consisting of 153 real-world scenarios; and propose \tool{}, a novel LLM-based deployability-centric framework that uses an iterative feedback mechanism to generate IaC templates.
Our evaluation reveals that state-of-the-art LLMs initially performed poorly, with Claude-3.5 and Claude-3.7 achieving only 30.2\% and 26.8\% deployment success on the first attempt, respectively. 
However, IaCGen transforms this performance dramatically: all evaluated models reach $>$90\% passItr@25, with Claude-3.5 and Claude-3.7 achieving 98\% success rate. 
This work provides the first comprehensive assessment of deployability-centric IaC template generation and establishes a foundation for future research.

\paratitle{Data Availability} The replication package contains two folders. The Data folder contains the benchmarks, and the Code folder contains the code of our \tool{} framework. Detailed descriptions of files can be found in the README.md file within each folder. The replication package is available at~\url{https://github.com/Tianyi2/IaCGen}.


\bibliographystyle{unsrtnat}
\bibliography{11_References}

@article{yang2022generating,
  title={Generating natural language proofs with verifier-guided search},
  author={Yang, Kaiyu and Deng, Jia and Chen, Danqi},
  journal={arXiv preprint arXiv:2205.12443},
  year={2022}
}

@inproceedings{coignion2024performance,
  title={A performance study of llm-generated code on leetcode},
  author={Coignion, Tristan and Quinton, Cl{\'e}ment and Rouvoy, Romain},
  booktitle={Proceedings of the 28th International Conference on Evaluation and Assessment in Software Engineering},
  pages={79--89},
  year={2024}
}

@inproceedings{sun2022mining,
  title={Mining android api usage to generate unit test cases for pinpointing compatibility issues},
  author={Sun, Xiaoyu and Chen, Xiao and Zhao, Yanjie and Liu, Pei and Grundy, John and Li, Li},
  booktitle={Proceedings of the 37th IEEE/ACM International Conference on Automated Software Engineering},
  pages={1--13},
  year={2022}
}

@article{sun2023taming,
  title={Taming android fragmentation through lightweight crowdsourced testing},
  author={Sun, Xiaoyu and Chen, Xiao and Liu, Yonghui and Grundy, John and Li, Li},
  journal={IEEE Transactions on Software Engineering},
  volume={49},
  number={6},
  pages={3599--3615},
  year={2023},
  publisher={IEEE}
}

@inproceedings{sun2023lazycow,
  title={LazyCow: A lightweight crowdsourced testing tool for taming Android fragmentation},
  author={Sun, Xiaoyu and Chen, Xiao and Liu, Yonghui and Grundy, John and Li, Li},
  booktitle={Proceedings of the 31st ACM Joint European Software Engineering Conference and Symposium on the Foundations of Software Engineering},
  pages={2127--2131},
  year={2023}
}

@article{sun2021taming,
  title={Taming reflection: An essential step toward whole-program analysis of android apps},
  author={Sun, Xiaoyu and Li, Li and Bissyand{\'e}, Tegawend{\'e} F and Klein, Jacques and Octeau, Damien and Grundy, John},
  journal={ACM Transactions on Software Engineering and Methodology (TOSEM)},
  volume={30},
  number={3},
  pages={1--36},
  year={2021},
  publisher={ACM New York, NY, USA}
}

@inproceedings{liu2022first,
  title={A first look at CI/CD adoptions in open-source android apps},
  author={Liu, Pei and Sun, Xiaoyu and Zhao, Yanjie and Liu, Yonghui and Grundy, John and Li, Li},
  booktitle={Proceedings of the 37th IEEE/ACM International Conference on Automated Software Engineering},
  pages={1--6},
  year={2022}
}

@article{zhou2025declarui,
  title={DeclarUI: Bridging Design and Development with Automated Declarative UI Code Generation},
  author={Zhou, Ting and Zhao, Yanjie and Hou, Xinyi and Sun, Xiaoyu and Chen, Kai and Wang, Haoyu},
  journal={Proceedings of the ACM on Software Engineering},
  volume={2},
  number={FSE},
  pages={219--241},
  year={2025},
  publisher={ACM New York, NY, USA}
}

@inproceedings{tao2025privacy,
  title={Privacy bills of materials (pribom): A transparent privacy information inventory for collaborative privacy notice generation in mobile app development},
  author={Tao, Zhen and Pan, Shidong and Xing, Zhenchang and Sun, Xiaoyu and Haggag, Omar and Grundy, John and Li, Jingjie and Zhu, Liming},
  booktitle={The 25th Privacy Enhancing Technologies Symposium},
  pages={392--409},
  year={2025},
  organization={Privacy Enhancing Technologies Board}
}

@article{gong2025towards,
  title={Towards Context-aware Mobile Privacy Notice: Implementation of A Deployable Contextual Privacy Policies Generator},
  author={Gong, Haochen and Tao, Zhen and Pan, Shidong and Xing, Zhenchang and Sun, Xiaoyu},
  journal={arXiv preprint arXiv:2509.22900},
  year={2025}
}

@inproceedings{li2024incremental,
  title={Incremental Context-free Grammar Inference in Black Box Settings},
  author={Li, Feifei and Chen, Xiao and Xiao, Xi and Sun, Xiaoyu and Chen, Chuan and Wang, Shaohua and Han, Jitao},
  booktitle={Proceedings of the 39th IEEE/ACM International Conference on Automated Software Engineering},
  pages={1171--1182},
  year={2024}
}

@article{liao2025navigating,
  title={Navigating the Labyrinth: Path-Sensitive Unit Test Generation with Large Language Models},
  author={Liao, Dianshu and Yin, Xin and Pan, Shidong and Ni, Chao and Xing, Zhenchang and Sun, Xiaoyu},
  journal={arXiv preprint arXiv:2509.23812},
  year={2025}
}

@inproceedings{koziolek2024llm,
  title={LLM-based and retrieval-augmented control code generation},
  author={Koziolek, Heiko and Gr{\"u}ner, Sten and Hark, Rhaban and Ashiwal, Virendra and Linsbauer, Sofia and Eskandani, Nafise},
  booktitle={Proceedings of the 1st International Workshop on Large Language Models for Code},
  pages={22--29},
  year={2024}
}

@inproceedings{gu2023llm,
  title={Llm-based code generation method for golang compiler testing},
  author={Gu, Qiuhan},
  booktitle={Proceedings of the 31st ACM Joint European Software Engineering Conference and Symposium on the Foundations of Software Engineering},
  pages={2201--2203},
  year={2023}
}

@article{mosqueira2023human,
  title={Human-in-the-loop machine learning: a state of the art},
  author={Mosqueira-Rey, Eduardo and Hern{\'a}ndez-Pereira, Elena and Alonso-R{\'\i}os, David and Bobes-Bascar{\'a}n, Jos{\'e} and Fern{\'a}ndez-Leal, {\'A}ngel},
  journal={Artificial Intelligence Review},
  volume={56},
  number={4},
  pages={3005--3054},
  year={2023},
  publisher={Springer}
}

@article{zanzotto2019human,
  title={Human-in-the-loop artificial intelligence},
  author={Zanzotto, Fabio Massimo},
  journal={Journal of Artificial Intelligence Research},
  volume={64},
  pages={243--252},
  year={2019}
}

@article{xiao2023llm,
  title={Llm a*: Human in the loop large language models enabled a* search for robotics},
  author={Xiao, Hengjia and Wang, Peng},
  journal={arXiv preprint arXiv:2312.01797},
  year={2023}
}

@article{cusumano2010cloud,
  title={Cloud computing and SaaS as new computing platforms},
  author={Cusumano, Michael},
  journal={Communications of the ACM},
  volume={53},
  number={4},
  pages={27--29},
  year={2010},
  publisher={ACM New York, NY, USA}
}

@inproceedings{qiu2023simplifying,
  title={Simplifying cloud management with cloudless computing},
  author={Qiu, Yiming and Kon, Patrick Tser Jern and Xing, Jiarong and Huang, Yibo and Liu, Hongyi and Wang, Xinyu and Huang, Peng and Chowdhury, Mosharaf and Chen, Ang},
  booktitle={Proceedings of the 22nd ACM Workshop on Hot Topics in Networks},
  pages={95--101},
  year={2023}
}

@misc{rackspace-report,
  author = {Rackspace},  
  title = {The 2025 State of Cloud Report},
  year = {2025}, 
  url = {https://www.rackspace.com/blog/2025-state-cloud-report},  
}

@inproceedings{nasiri2024towards,
  title={Towards a Taxonomy of Infrastructure as Code Misconfigurations: An Ansible Study},
  author={Nasiri, Roya and Kumara, Indika and Tamburri, Damian Andrew and van den Heuvel, Willem-Jan},
  booktitle={Symposium and Summer School on Service-Oriented Computing},
  pages={83--103},
  year={2024},
  organization={Springer}
}

@article{kumara2021s,
  title={The do’s and don’ts of infrastructure code: A systematic gray literature review},
  author={Kumara, Indika and Garriga, Mart{\'\i}n and Romeu, Angel Urbano and Di Nucci, Dario and Palomba, Fabio and Tamburri, Damian Andrew and van den Heuvel, Willem-Jan},
  journal={Information and Software Technology},
  volume={137},
  pages={106593},
  year={2021},
  publisher={Elsevier}
}

@misc{basher2019devops,
  title={DevOps: An explorative case study on the challenges and opportunities in implementing Infrastructure as code},
  author={Basher, Mohamed},
  year={2019}
}

@misc{hashicorp2023,
  author = {Hashicorp},
  title = {Hashicorp 2023 state of cloud strategy survey.},
  year = {2023},
  url = {https://www.hashicorp.com/en/state-of-the-cloud},
}

@misc{iac_low,
  author = {Styra},
  title  = {AI-Generated Infrastructure-as-Code: The Good, the Bad and the Ugly},
  year   = {2023},
  note   = {\url{https://direct-blog-url-here.com}}
}

@phdthesis{cirlan2024mining,
  title={Mining for Cost Awareness in Cloud Computing: A Study of AWS CloudFormation and Developer Practices},
  author={Cirlan, Alexandru-Nicolae},
  school={University of Groningen},
  year={2024}
}

@misc{aws_cfn_templates,
  note = {aws-cloudformation-templates by AWS Cloudformation : AWS CloudFormation Sample Templates. \url{https://github.com/aws-cloudformation/aws-cloudformation-templates}}
}

@misc{aws_samples_github,
  author = {AWS Samples},
  title = {AWS Samples GitHub Repository},
  year = {2025},
  howpublished = {\url{https://github.com/aws-samples}},
}

@article{lachaux2020unsupervised,
  title={Unsupervised translation of programming languages},
  author={Lachaux, Marie-Anne and Roziere, Baptiste and Chanussot, Lowik and Lample, Guillaume},
  journal={arXiv preprint arXiv:2006.03511},
  year={2020}
}

@article{roziere2021leveraging,
  title={Leveraging automated unit tests for unsupervised code translation},
  author={Roziere, Baptiste and Zhang, Jie M and Charton, Francois and Harman, Mark and Synnaeve, Gabriel and Lample, Guillaume},
  journal={arXiv preprint arXiv:2110.06773},
  year={2021}
}

@article{austin2021program,
  title={Program synthesis with large language models},
  author={Austin, Jacob and Odena, Augustus and Nye, Maxwell and Bosma, Maarten and Michalewski, Henryk and Dohan, David and Jiang, Ellen and Cai, Carrie and Terry, Michael and Le, Quoc and others},
  journal={arXiv preprint arXiv:2108.07732},
  year={2021}
}

@article{kon2024iac,
  title={IaC-Eval: A Code Generation Benchmark for Cloud Infrastructure-as-Code Programs},
  author={Kon, Patrick T and Liu, Jiachen and Qiu, Yiming and Fan, Weijun and He, Ting and Lin, Lei and Zhang, Haoran and Park, Owen M and Elengikal, George S and Kang, Yuxin and others},
  journal={Advances in Neural Information Processing Systems},
  volume={37},
  pages={134488--134506},
  year={2024}
}

@article{madaan2023self,
  title={Self-refine: Iterative refinement with self-feedback},
  author={Madaan, Aman and Tandon, Niket and Gupta, Prakhar and Hallinan, Skyler and Gao, Luyu and Wiegreffe, Sarah and Alon, Uri and Dziri, Nouha and Prabhumoye, Shrimai and Yang, Yiming and others},
  journal={Advances in Neural Information Processing Systems},
  volume={36},
  pages={46534--46594},
  year={2023}
}

@article{chen2023teaching,
  title={Teaching large language models to self-debug},
  author={Chen, Xinyun and Lin, Maxwell and Sch{\"a}rli, Nathanael and Zhou, Denny},
  journal={arXiv preprint arXiv:2304.05128},
  year={2023}
}

@article{huang2022large,
  title={Large language models can self-improve},
  author={Huang, Jiaxin and Gu, Shixiang Shane and Hou, Le and Wu, Yuexin and Wang, Xuezhi and Yu, Hongkun and Han, Jiawei},
  journal={arXiv preprint arXiv:2210.11610},
  year={2022}
}

@article{chinamanagonda2019automating,
  title={Automating Infrastructure with Infrastructure as Code (IaC)},
  author={Chinamanagonda, Sandeep},
  journal={Available at SSRN 4986767},
  year={2019}
}

@article{palavalli2024using,
  title={Using a Feedback Loop for LLM-based Infrastructure as Code Generation},
  author={Palavalli, Mayur Amarnath and Santolucito, Mark},
  journal={arXiv preprint arXiv:2411.19043},
  year={2024}
}

@inproceedings{ragothaman2024optimizing,
  title={Optimizing Service Deployments With NLP Based Infrastructure Code Generation-An Automation Framework},
  author={Ragothaman, Hariharan and Udayakumar, Saai Krishnan},
  booktitle={2024 IEEE 2nd International Conference on Electrical Engineering, Computer and Information Technology (ICEECIT)},
  pages={216--221},
  year={2024},
  organization={IEEE}
}

@article{chen2021evaluating,
  title={Evaluating large language models trained on code},
  author={Chen, Mark and Tworek, Jerry and Jun, Heewoo and Yuan, Qiming and Pinto, Henrique Ponde De Oliveira and Kaplan, Jared and Edwards, Harri and Burda, Yuri and Joseph, Nicholas and Brockman, Greg and others},
  journal={arXiv preprint arXiv:2107.03374},
  year={2021}
}

@misc{cfn-linter,
  author = {Kevin, DeJong and Pat, Myron and Frank, van Boven and others},
  title = {AWS CloudFormation Linter},
  year = {2025},
  url = {https://github.com/aws-cloudformation/cfn-lint},
}

@misc{yaml-linter,
  author = {Adrien Vergé et al.},
  title = {yamllint},
  year = {2023},
  url = {https://github.com/adrienverge/yamllint},
}

@article{drosos2024your,
  title={When your infrastructure is a buggy program: Understanding faults in infrastructure as code ecosystems},
  author={Drosos, Georgios-Petros and Sotiropoulos, Thodoris and Alexopoulos, Georgios and Mitropoulos, Dimitris and Su, Zhendong},
  journal={Proceedings of the ACM on Programming Languages},
  volume={8},
  number={OOPSLA2},
  pages={2490--2520},
  year={2024},
  publisher={ACM New York, NY, USA}
}

@misc{checkov,
  author = {Nimrod, Kor and Anton, Grübel and James, Woolfenden and others},
  title = {Checkov},
  year = {2025},
  url = {https://github.com/bridgecrewio/checkov},
}

@inproceedings{sokolowski2022infrastructure,
  title={Infrastructure as code for dynamic deployments},
  author={Sokolowski, Daniel},
  booktitle={Proceedings of the 30th ACM Joint European Software Engineering Conference and Symposium on the Foundations of Software Engineering},
  pages={1775--1779},
  year={2022}
}

@misc{iac-best-practices,
  author = {Vahid Iranpour},
  title = {Mastering Infrastructure as Code (IaC): Best Practices and Real-World Examples},
  year = {2024},
  url = {https://medium.com/@community.vahid/mastering-infrastructure-as-code-iac-best-practices-and-real-world-examples-df0f3c90c560},
}

@inproceedings{hasan2020testing,
  title={Testing practices for infrastructure as code},
  author={Hasan, Mohammed Mehedi and Bhuiyan, Farzana Ahamed and Rahman, Akond},
  booktitle={Proceedings of the 1st ACM SIGSOFT International Workshop on Languages and Tools for Next-Generation Testing},
  pages={7--12},
  year={2020}
}

@inproceedings{guerriero2019adoption,
  title={Adoption, support, and challenges of infrastructure-as-code: Insights from industry},
  author={Guerriero, Michele and Garriga, Martin and Tamburri, Damian A and Palomba, Fabio},
  booktitle={2019 IEEE International conference on software maintenance and evolution (ICSME)},
  pages={580--589},
  year={2019},
  organization={IEEE}
}

@article{vessey1985expertise,
  title={Expertise in debugging computer programs: A process analysis},
  author={Vessey, Iris},
  journal={International Journal of Man-Machine Studies},
  volume={23},
  number={5},
  pages={459--494},
  year={1985},
  publisher={Elsevier}
}

@misc{debugging-web,
  author = {James saloman},
  title = {Debugging 101: How to Find and Fix Programming Errors},
  year = {2023},
  url = {https://dev.to/saloman_james/debugging-101-how-to-find-and-fix-programming-errors-1d14},
}

@misc{tf-limitation-first,
  author = {Sean},
  title = {How Terraform manage apply failure},
  year = {2022},
  url = {https://stackoverflow.com/questions/72070394/how-terraform-manage-apply-failure},
}

@misc{tf-limitation-second,
  author = {HashiCorp},
  title = {Apply previous state file or undo plan},
  year = {2023},
  url = {https://discuss.hashicorp.com/t/apply-previous-state-file-or-undo-plan/57833},
}

@misc{cf-stack,
  author = {AWS},
  title = {Managing AWS resources as a single unit with AWS CloudFormation stacks},
  year = {2025},
  url = {https://docs.aws.amazon.com/AWSCloudFormation/latest/UserGuide/stacks.html},
}

@article{liu2023your,
  title={Is your code generated by chatgpt really correct? rigorous evaluation of large language models for code generation},
  author={Liu, Jiawei and Xia, Chunqiu Steven and Wang, Yuyao and Zhang, Lingming},
  journal={Advances in Neural Information Processing Systems},
  volume={36},
  pages={21558--21572},
  year={2023}
}

@article{sobo2025evaluating,
  title={Evaluating LLMs for code generation in HRI: A comparative study of ChatGPT, gemini, and claude},
  author={Sobo, Andrei and Mubarak, Awes and Baimagambetov, Almas and Polatidis, Nikolaos},
  journal={Applied Artificial Intelligence},
  volume={39},
  number={1},
  pages={2439610},
  year={2025},
  publisher={Taylor \& Francis}
}

@article{islam2025codesim,
  title={CODESIM: Multi-Agent Code Generation and Problem Solving through Simulation-Driven Planning and Debugging},
  author={Islam, Md Ashraful and Ali, Mohammed Eunus and Parvez, Md Rizwan},
  journal={arXiv preprint arXiv:2502.05664},
  year={2025}
}

@article{guo2024deepseek,
  title={DeepSeek-Coder: When the Large Language Model Meets Programming--The Rise of Code Intelligence},
  author={Guo, Daya and Zhu, Qihao and Yang, Dejian and Xie, Zhenda and Dong, Kai and Zhang, Wentao and Chen, Guanting and Bi, Xiao and Wu, Yu and Li, YK and others},
  journal={arXiv preprint arXiv:2401.14196},
  year={2024}
}

@article{pan2024textit,
  title={"I Don't Use AI for Everything": Exploring Utility, Attitude, and Responsibility of AI-empowered Tools in Software Development},
  author={Pan, Shidong and Wang, Litian and Zhang, Tianyi and Xing, Zhenchang and Zhao, Yanjie and Lu, Qinghua and Sun, Xiaoyu},
  journal={arXiv preprint arXiv:2409.13343},
  year={2024}
}

@misc{deepseek-share,
  author = {Naveen Kumar},
  title = {DeepSeek AI Statistics of 2025 (Users and Revenue)},
  year = {2025},
  url = {https://www.demandsage.com/deepseek-statistics/},
}

@inproceedings{jain2023skyplane,
  title={Skyplane: Optimizing transfer cost and throughput using $\{$Cloud-Aware$\}$ overlays},
  author={Jain, Paras and Kumar, Sam and Wooders, Sarah and Patil, Shishir G and Gonzalez, Joseph E and Stoica, Ion},
  booktitle={20th USENIX Symposium on Networked Systems Design and Implementation (NSDI 23)},
  pages={1375--1389},
  year={2023}
}

@inproceedings{stoica2021cloud,
  title={From cloud computing to sky computing},
  author={Stoica, Ion and Shenker, Scott},
  booktitle={Proceedings of the Workshop on Hot Topics in Operating Systems},
  pages={26--32},
  year={2021}
}

@article{pum2024cloudformation,
  title={CloudFormation vs. Terraform: A Comparative Study},
  author={Pum, Mengkorn and George, Amelia},
  year={2024}
}

@misc{hash-tf-cf-compare,
  author = {Hashicorp},
  title = {Terraform versus CloudFormation, Heat, and other infrastructure as code tools},
  year = {2024},
  url = {https://developer.hashicorp.com/terraform/intro/vs/cloudformation},
}

@misc{geeks-tf-cf-compare,
  author = {GeeksForGeeks},
  title = {Difference Between CloudFormation VS Terraform},
  year = {2024},
  url = {https://www.geeksforgeeks.org/difference-between-cloudformation-vs-terraform/},
}

@misc{aws-doc,
  author = {AWS},
  title = {AWS CloudFormation User Guide},
  year = {2025},
  url = {https://docs.aws.amazon.com/AWSCloudFormation/latest/UserGuide/Welcome.html},
}

@misc{aws-best-practices,
  author = {AWS},
  title = {AWS Prescriptive Guidance},
  year = {2025},
  url = {https://docs.aws.amazon.com/prescriptive-guidance/latest/iac-edp-combo-approach/best-practices.html},
}

@article{liao2024code,
  title={A 3-CodGen: A Repository-Level Code Generation Framework for Code Reuse with Local-Aware, Global-Aware, and Third-Party-Library-Aware},
  author={Liao, Dianshu and Pan, Shidong and Sun, Xiaoyu and Ren, Xiaoxue and Huang, Qing and Xing, Zhenchang and Jin, Huan and Li, Qinying},
  journal={IEEE Transactions on Software Engineering},
  year={2024},
  publisher={IEEE}
}

@article{han2024chase,
  title={Do Chase Your Tail! Missing Key Aspects Augmentation in Textual Vulnerability Descriptions of Long-tail Software through Feature Inference},
  author={Han, Linyi and Pan, Shidong and Xing, Zhenchang and Sun, Jiamou and Yitagesu, Sofonias and Zhang, Xiaowang and Feng, Zhiyong},
  journal={IEEE Transactions on Software Engineering},
  year={2024},
  publisher={IEEE}
}

@article{si2024solution,
  title={A solution toward transparent and practical AI regulation: Privacy nutrition labels for open-source generative AI-based applications},
  author={Si, Meixue and Pan, Shidong and Liao, Dianshu and Sun, Xiaoyu and Tao, Zhen and Shi, Wenchang and Xing, Zhenchang},
  journal={arXiv preprint arXiv:2407.15407},
  year={2024}
}

@article{zhou2024bridging,
  title={Bridging design and development with automated declarative ui code generation},
  author={Zhou, Ting and Zhao, Yanjie and Hou, Xinyi and Sun, Xiaoyu and Chen, Kai and Wang, Haoyu},
  journal={arXiv preprint arXiv:2409.11667},
  year={2024}
}

@article{pan2024large,
  title={A Large-scale Investigation of Semantically Incompatible APIs behind Compatibility Issues in Android Apps},
  author={Pan, Shidong and Guo, Tianchen and Zhang, Lihong and Liu, Pei and Xing, Zhenchang and Sun, Xiaoyu},
  journal={arXiv preprint arXiv:2406.17431},
  year={2024}
}

@article{bi2024iterative,
  title={Iterative refinement of project-level code context for precise code generation with compiler feedback},
  author={Bi, Zhangqian and Wan, Yao and Wang, Zheng and Zhang, Hongyu and Guan, Batu and Lu, Fangxin and Zhang, Zili and Sui, Yulei and Jin, Hai and Shi, Xuanhua},
  journal={arXiv preprint arXiv:2403.16792},
  year={2024}
}

@article{bouzenia2025you,
  title={You name it, I run it: An LLM agent to execute tests of arbitrary projects},
  author={Bouzenia, Islem and Pradel, Michael},
  journal={Proceedings of the ACM on Software Engineering},
  volume={2},
  number={ISSTA},
  pages={1054--1076},
  year={2025},
  publisher={ACM New York, NY, USA}
}

@article{zhang2023repocoder,
  title={Repocoder: Repository-level code completion through iterative retrieval and generation},
  author={Zhang, Fengji and Chen, Bei and Zhang, Yue and Keung, Jacky and Liu, Jin and Zan, Daoguang and Mao, Yi and Lou, Jian-Guang and Chen, Weizhu},
  journal={arXiv preprint arXiv:2303.12570},
  year={2023}
}

@article{dong2022survey,
  title={A survey on in-context learning},
  author={Dong, Qingxiu and Li, Lei and Dai, Damai and Zheng, Ce and Ma, Jingyuan and Li, Rui and Xia, Heming and Xu, Jingjing and Wu, Zhiyong and Liu, Tianyu and others},
  journal={arXiv preprint arXiv:2301.00234},
  year={2022}
}

@misc{cis-aws,
  author       = {Center for Internet Security},
  title        = {CIS AWS Foundations Benchmark, v5.0.0},
  year         = {2025},
  url = {https://www.cisecurity.org/benchmark/amazon_web_services},
}

@misc{cis,
  author       = {Center for Internet Security},
  title        = {CIS Benchmarks for Cloud Providers},
  year         = {n.d.},
  url = {https://www.cisecurity.org/cis-benchmarks},
}

@inproceedings{pujar2023automated,
  title={Automated code generation for information technology tasks in yaml through large language models},
  author={Pujar, Saurabh and Buratti, Luca and Guo, Xiaojie and Dupuis, Nicolas and Lewis, Burn and Suneja, Sahil and Sood, Atin and Nalawade, Ganesh and Jones, Matt and Morari, Alessandro and others},
  booktitle={2023 60th ACM/IEEE Design Automation Conference (DAC)},
  pages={1--4},
  year={2023},
  organization={IEEE}
}

@article{srivatsa2024survey,
  title={A survey of using large language models for generating infrastructure as code},
  author={Srivatsa, Kalahasti Ganesh and Mukhopadhyay, Sabyasachi and Katrapati, Ganesh and Shrivastava, Manish},
  journal={arXiv preprint arXiv:2404.00227},
  year={2024}
}

@misc{iac-ci-cd,
  author = {Microsoft Azure},
  title = {Architecture strategies for using infrastructure as code},
  year = {2023},
  url = {https://learn.microsoft.com/en-us/azure/well-architected/operational-excellence/infrastructure-as-code-design},
}

@article{minna2025analyzing,
  title={Analyzing and mitigating (with llms) the security misconfigurations of helm charts from artifact hub},
  author={Minna, Francesco and Massacci, Fabio and Tuma, Katja},
  journal={Empirical Software Engineering},
  volume={30},
  number={5},
  pages={132},
  year={2025},
  publisher={Springer}
}

@misc{bucket_naming_solution,
  author = {Vijay Kuma},  
  title = {S3 Bucket Naming Conventions: Best Practices to Avoid Unexpected Costs},
  year = {2024}, 
  url = {https://tech.vians.org/posts/s3-bucket-naming-conventions/},  
}

@misc{bucket_naming_guide,
  author = {AWS},  
  title = {General purpose bucket naming rules},
  year = {n.d.}, 
  url = {https://docs.aws.amazon.com/AmazonS3/latest/userguide/bucketnamingrules.html},  
}

@misc{ami_solution,
  author = {Martin Yip},  
  title = {Query for the latest Amazon Linux AMI IDs using AWS Systems Manager Parameter Store},
  year = {2018}, 
  url = {https://aws.amazon.com/blogs/compute/query-for-the-latest-amazon-linux-ami-ids-using-aws-systems-manager-parameter-store/},  
}

@misc{dbinstance_solution,
  author = {AWS},  
  title = {CreateDBInstanceReadReplica},
  year = {n.d.}, 
  url = {https://docs.aws.amazon.com/AmazonRDS/latest/APIReference/API_CreateDBInstanceReadReplica.html}}

@misc{stackoverflow,
  author = {Stack Overflow},  
  title = {Stack Overflow website},
  year = {n.d.}, 
  url = {https://stackoverflow.com/questions},  
}

@misc{prisma_cloud,
  author = {Prisma Cloud},  
  title = {Prisma Cloud Website},
  year = {n.d.}, 
  url = {https://www.paloaltonetworks.com/prisma/cloud},  
}

@misc{low_severity_policy,
  author = {Prisma Cloud},  
  title = {AWS security groups allow ingress from 0.0.0.0/0 to port 80},
  year = {2025}, 
  url = {https://docs.prismacloud.io/en/enterprise-edition/policy-reference/aws-policies/aws-networking-policies/ensure-aws-security-groups-do-not-allow-ingress-from-00000-to-port-80},  
}

@misc{medium_severity_policy,
  author = {Prisma Cloud},  
  title = {AWS SNS topic has SSE disabled},
  year = {2025}, 
  url = {https://docs.prismacloud.io/en/enterprise-edition/policy-reference/aws-policies/aws-general-policies/general-15},  
}

@inproceedings{mathew2014study,
  title={A study of open ports as security vulnerabilities in common user computers},
  author={Mathew, Kuruvilla and Tabassum, Mujahid and Siok, Marlene Valerie Lu Ai},
  booktitle={2014 International Conference on Computational Science and Technology (ICCST)},
  pages={1--6},
  year={2014},
  organization={IEEE}
}

@article{dahabiyeh2021ignorance,
  title={When ignorance is bliss: The role of curiosity in online games adoption},
  author={Dahabiyeh, Laila and Najjar, Mohammad S and Agrawal, Deepti},
  journal={Entertainment Computing},
  volume={37},
  pages={100398},
  year={2021},
  publisher={Elsevier}
}

@misc{low_severity_impact,
  author = {Jonathan Carpenter},  
  title = {Why Ignoring Low-Severity Vulnerabilities Can Cost You: A Risk-Based Patch Management Guide: Part Four},
  year = {2024}, 
  url = {https://anchorcybersecurity.com/blog/2024-10-25-Cost-of-Ignoring},  
}

@misc{public_problematic_template_SNS,
  author = {Eric Z. Beard},  
  title = {AWS CloudFormation Sample Templates},
  year = {2024}, 
  url = {https://github.com/aws-cloudformation/aws-cloudformation-templates/blob/main/SNS/SNSTopic.yaml},  
}

@misc{public_problematic_template_security_group,
  author = {Luis Casas},  
  title = {aws projects},
  year = {2025}, 
  url = {https://github.com/LSCasas/aws_projects/blob/a19f570a682028186078ff7396e
58a2e9f6c612e/ecs-fargate/CloudFormation_Template/.yaml#L105},  
}

@misc{checkov-custom-policy,
  author = {Checkov},  
  title = {Custom Policies Overview},
  year = {n.d.}, 
  url = {https://www.checkov.io/3.Custom%20Policies/Custom%20Policies%20Overview.html},  
}

@article{liu2025ai,
  title={When AI Takes the Wheel: Security Analysis of Framework-Constrained Program Generation},
  author={Liu, Yue and Xing, Zhenchang and Pan, Shidong and Tantithamthavorn, Chakkrit},
  journal={arXiv preprint arXiv:2510.16823},
  year={2025}
}

\end{document}